\documentclass[prd,twocolumn,nofootinbib,preprint,superscriptaddress,10pt]{revtex4}
\pdfoutput=1
\usepackage[utf8]{inputenc}
\usepackage{subfigure}
\usepackage{epsfig} 
\usepackage{amsmath} 
\usepackage{amsfonts}  
\usepackage{float}    
\usepackage{amssymb}   
\usepackage{slashed} 
\usepackage{color}       
\usepackage{pbox}
\usepackage{array,multirow,makecell}
\usepackage[colorlinks,citecolor=blue, linkcolor = blue, urlcolor = blue]{hyperref}
\usepackage{tabularx}
\usepackage{titlesec} 
\usepackage{scrextend}
\usepackage{ulem}
\usepackage{natbib} 
\usepackage{array,multirow,makecell}
\usepackage{lipsum}
\newcommand{\be}{\begin{equation}}
\newcommand{\ee}{\end{equation}}
\newcommand{\bea}{\begin{eqnarray}}
\newcommand{\eea}{\end{eqnarray}}
\newcommand{\bew}{\begin{widetext}}
\newcommand{\eew}{\end{widetext}}
\begin{document}
%========================================================
%\title{Note on Fermion doublet dark matter}
\title{Reviving sub-TeV $SU(2)_L$ lepton doublet Dark Matter}
%=========================
\author{Purusottam Ghosh}
\email{spspg2655@iacs.res.in}
\affiliation{School of Physical Sciences, Indian Association for the Cultivation of Science,\\ 2A $\&$ 2B Raja S.C. Mullick Road, Kolkata 700032, India}

\author{Sk Jeesun}
\email{skjeesun48@gmail.com}
\affiliation{School of Physical Sciences, Indian Association for the Cultivation of Science,\\ 2A $\&$ 2B Raja S.C. Mullick Road, Kolkata 700032, India}
%======================================================================== 
\begin{abstract}
In this work we study the hybrid kind of dark matter(DM) production mechanism where both thermal and non-thermal contribution at two different epochs set the DM relic abundance. This hybrid set up in turn shifts the parameter space of DM in contrast to pure thermal DM scenario.
We review such production mechanism in the context of the $SU(2)_L$ lepton doublet dark matter ($\Psi$) augmented with an additional singlet dark scalar ($S$). The neutral component of the dark doublet can serve as a stable DM candidate and in pure thermal scenario, it is under-abundant as well as excluded from direct detection constraints due to its strong gauge interactions in the sub-TeV mass regime. However, in addition to the thermal contribution, the late time non-thermal DM production from the decay of the long-lived dark scalar $S$ helps to fulfill the deficit in DM abundance. On the other hand, the strong gauge mediated direct detection constraint can be evaded with the help of a $SU(2)_L$ triplet scalar(with $Y=2$), resulting a pseudo-Dirac DM. 
To realize our proposed scenario we impose a discrete $\mathcal{Z}_2$ symmetry under which both $\Psi$ and $S$ are odd while rest of the fields are even.
We find the lepton doublet pseudo-Dirac DM with mass $\sim 450-1200$ GeV, compatible with the observed relic density, direct, indirect, and existing collider search constraints.

\end{abstract}
\maketitle
%%%%%%%%%%%%%%%%%%%%%%%%%%%%%%%%%
\section{Introduction}
%=================================
The existence of Dark matter (DM) is strongly suggested by several
astrophysical and cosmological observations at a wide range of length scales concluding that about 80-85$\%$ of total matter density is dominated by DM \cite{Zwicky:1933gu,Rubin:1970zza,Clowe:2006eq,Planck:2018vyg}.
In spite of this great 
observational evidence, the Standard Model(SM) of particle physics at  present set up fails to  explain the particle DM.
Additionally, the SM also can not explain the existence of neutrino mass and mixing as
suggested by different neutrino oscillation experiments\cite{T2K:2011ypd,
DoubleChooz:2011ymz,DayaBay:2012fng,RENO:2012mkc,MINOS:2013xrl,
ParticleDataGroup:2020ssz}.
To overcome these two shortcomings of the SM,  various beyond standard model(BSM) scenarios have been proposed.
The issue of neutrino masses and their mixing angles can be resolved by the three seesaw
mechanisms \cite{Minkowski:1977sc,Mohapatra:1979ia,
Schechter:1980gr,Gell-Mann:1979vob, Mohapatra:1980yp,Lazarides:1980nt,
Wetterich:1981bx,Schechter:1981cv,Brahmachari:1997cq,Foot:1988aq}.
Weakly interacting massive particles (WIMP) \cite{Kolb:1990vq,Feng:2010gw,Roszkowski:2017nbc,Schumann:2019eaa,
Lin:2019uvt,Arcadi:2017kky} is the most popular and widely studied thermal DM candidate whose interaction strength with SM particles is assumed to be of the order of electroweak interactions to explain the observed relic density. 
However, the null results at various direct detection experiments \cite{PandaX-II:2016vec, XENON:2017vdw, LUX:2016ggv,PICO:2019vsc,
AMS:2013fma,Buckley:2013bha,Gaskins:2016cha,Fermi-LAT:2016uux,MAGIC:2016xys,
Bringmann:2012ez,Cirelli:2015gux,Kahlhoefer:2017dnp,Boveia:2018yeb} open the possibilities of alternate paradigms to explain the DM relic density. Among the alternate theories, the feebly interacting massive particle(FIMP) is a very attractive viable candidate whose interactions with SM plasma are considered to be too small to keep them in a thermal bath \cite{Hall:2009bx,Konig:2016dzg,
Biswas:2016bfo, Bernal:2017kxu,Borah:2018gjk,Ghosh:2023ocl}.
Rather FIMPs are produced non-thermally from the decay or annihilation of bath particles and  number density gradually freezes in and such a scenario is known as non-thermal DM scenario. 
Such tiny
interactions of FIMP with SM particles can be the possible reason for the non-observations of DM in different detection experiments such
as Panda\cite{PandaX-II:2016vec}, XENON\cite{XENON:2017vdw}, LUX\cite{LUX:2016ggv}.

Despite the fact that the dedicated direct search experiments have put the thermal WIMP paradigm in a corner and non-thermal DM scenarios are being 
explored widely as an alternate theory, yet there are several approaches to 
revive thermal DM scenario itself \cite{Hochberg:2014dra,DEramo:2017gpl,Medina:2014bga,Puetter:2022ucx,Frumkin:2021zng,Fairbairn:2008fb,DiazSaez:2023wli}.
Examples include strongly interacting massive particle(SIMP) \cite{Hochberg:2014dra}, assuming non-standard cosmology \cite{DEramo:2017gpl}, bouncing dark matter \cite{Puetter:2022ucx}, freeze out from inverse decays \cite{Frumkin:2021zng}. 
Among the different methods to resurrect thermal DM scenarios with different phenomenological implications,
we are interested in a hybrid scenario where both the thermal and non-thermal contributions set the DM relic abundance.
In such cases, the DM candidate can have sizeable interaction with SM bath particles giving rise to suppressed number density but non-thermal production from another source helps to meet observed relic giving rise to different observational consequences \cite{Fairbairn:2008fb,Borah:2017dfn,Biswas:2018ybc}.
At this point, it is worth mentioning that there exist some well-motivated particle physics models which can account for single component thermal DM but suffers from the under-abundance issue in a certain mass range i.e. can not 
explain  $100\%$ of the observed relic density \cite{Ma:2006km,Ma:2008cu, Arkani-Hamed:2005zuc,DEramo:2007anh}. 
And for such models, the earlier mentioned hybrid scenario is beneficial where apart from the thermal density additional contribution is required to satisfy the whole fraction of DM relic density.
In this work to realize such a mechanism we target the vector-like $SU(2)_L$ 
lepton doublet dark matter model \cite{DEramo:2007anh}. 
%in addition to SM particle content \cite{Bhattacharya:2018fus}.
The striking feature of such a model is that it has only one free parameter, the DM mass. The interaction couplings with SM particles are gauge couplings and are not free parameters. 
However, in such a scenario, the strong gauge-mediated interaction of DM with SM bath leads to suppressed number density making it under abundant for masses below $1.2$ TeV \cite{Bhattacharya:2018fus} \footnote{On the contrary pure thermal fermion DM has also been studied in presence of an additional singlet fermion \cite{Mahbubani:2005pt,Bhattacharya:2015qpa,Bhattacharya:2017sml,Bhattacharya:2018fus,Barman:2019aku,Ghosh:2021wrk,Dey:2022whc,Bhattacharya:2021ltd,Konar:2020wvl,Konar:2020vuu,Bhattacharya:2022wtr}.}. 
We are interested in the sub-TeV range of $SU(2)_L$  lepton doublet dark matter and revisit such setup whether non-thermal contribution from additional sources can elevate the under-abundant region.

As mentioned above, our aim  in this work is to study the phenomenology of a vector like $SU(2)_L$  lepton doublet, $\Psi=(\psi^0~~\psi^-)^T$ where the neutral component, $\psi^0$ serves the role of the dark matter. In order to realize sub-TeV lepton doublet type DM we extend the dark sector with another dark real scalar, $S$. 
The stability of the lightest dark sector particle, $\psi^0$ which acts as DM can be ensured with the help of additional discrete symmetry, $\mathcal{Z}_2$, under which both the fields ($\Psi$,$S$) are odd and the rest of the particles are even.
The scalar $S$ has Higgs portal coupling with SM particles due to which $S$
freezes out from the thermal bath.
On the other hand, $\psi^0$ also freezes out from the bath decided by all number-changing processes including co-annihilation.
Due to the chosen parameter space $S$ freezes out at an earlier time than the time when $\psi^0$ freezes out.
However, $S$ has Yukawa coupling with $\psi^0$ and for $M_S> M_{\psi^0}$, $S$ can decay to $\psi^0$ increasing $\psi^0$ number density.
To repopulate $\psi^0$ number density from the non-thermal late decay of $S$ the Yukawa coupling of $S$ with  $\psi^0$ should be $\lesssim 10^{-9}$.
%We demand the Yukawa coupling of the interaction between $\psi^0$ and $S$ to be tiny for non thermal production 
And throughout our analysis we choose it to be $10^{-10}$ for not tampering with the prediction of light abundances from Big Bang nucleosynthesis (BBN)\cite{Nollett:2014lwa}.
We discuss all the observational constraints like direct detection, indirect detection, and collider search for our setup. 

The remaining part of this paper is organized as follows. First, we discuss the general framework of the DM production mechanism in Section \ref{sec:gen}. 
In Section \ref{sec:diracdm}, we study the phenomenology of  Dirac lepton doublet DM model. 
We then present a detailed discussion of the pseudo-Dirac lepton doublet DM, including DM abundance, direct, indirect, and collider search constraints in Section \ref{sec:psudodirac}. Finally, we conclude in Section\ref{sec:conc}. We also include some analysis relevant for our discussion in Appendix \ref{sec:apxdirac}-\ref{sec:loop}. 

%%%%%%%%%%%%%%%%%%%%%%%%%%%%%%%%%%%%%%%%%%%%%%%
\section{A General Framework of hybrid set up}
\label{sec:gen}
%===========================
%=====================
%\bew
\begin{figure}[!tbh]
    \centering
    \includegraphics[scale=0.3]{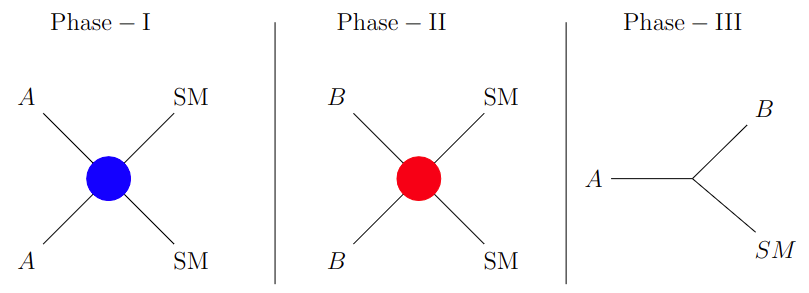}
    \caption{\it Cartoon diagrams of relevant interactions active during three different epochs of the Universe to set the relic abundance.}
    \label{fig:cartoon_int}
\end{figure}
%\eew
%=======================
In this work, our goal is to investigate the under-abundant region of the $SU(2)_L$ lepton doublet DM via late-time DM generation from a heavy thermal dark sector particle.
To alleviate the under-abundance issue we discuss here a hybrid type of DM
production mechanism with the help of another heavy dark sector particle which was in also thermal equilibrium at the early time of the Universe and yields non-zero number density via the freeze-out mechanism. 
Before going to the relic density analysis for specific $SU(2)_L$ lepton doublet DM we portray here a general discussion of the hybrid DM production mechanism \footnote{The general framework can be
applicable for any $SU(2)_L$ scalar or fermion multiplets $(n \geq 2)$ (often named
as electroweak DM) where DM remains underabundant in certain mass region
due to strong gauge interactions\cite{Ma:2006km,Ma:2008cu, Arkani-Hamed:2005zuc,DEramo:2007anh}. This can also be applicable for Majorana type DM scenarios where DM faces under-abundance issues as outlined in refs.\cite{Mahanta:2019gfe, Bonilla:2019ipe}.}. 

\begin{figure}[!tbh]
    \centering
    \includegraphics[scale=0.32]{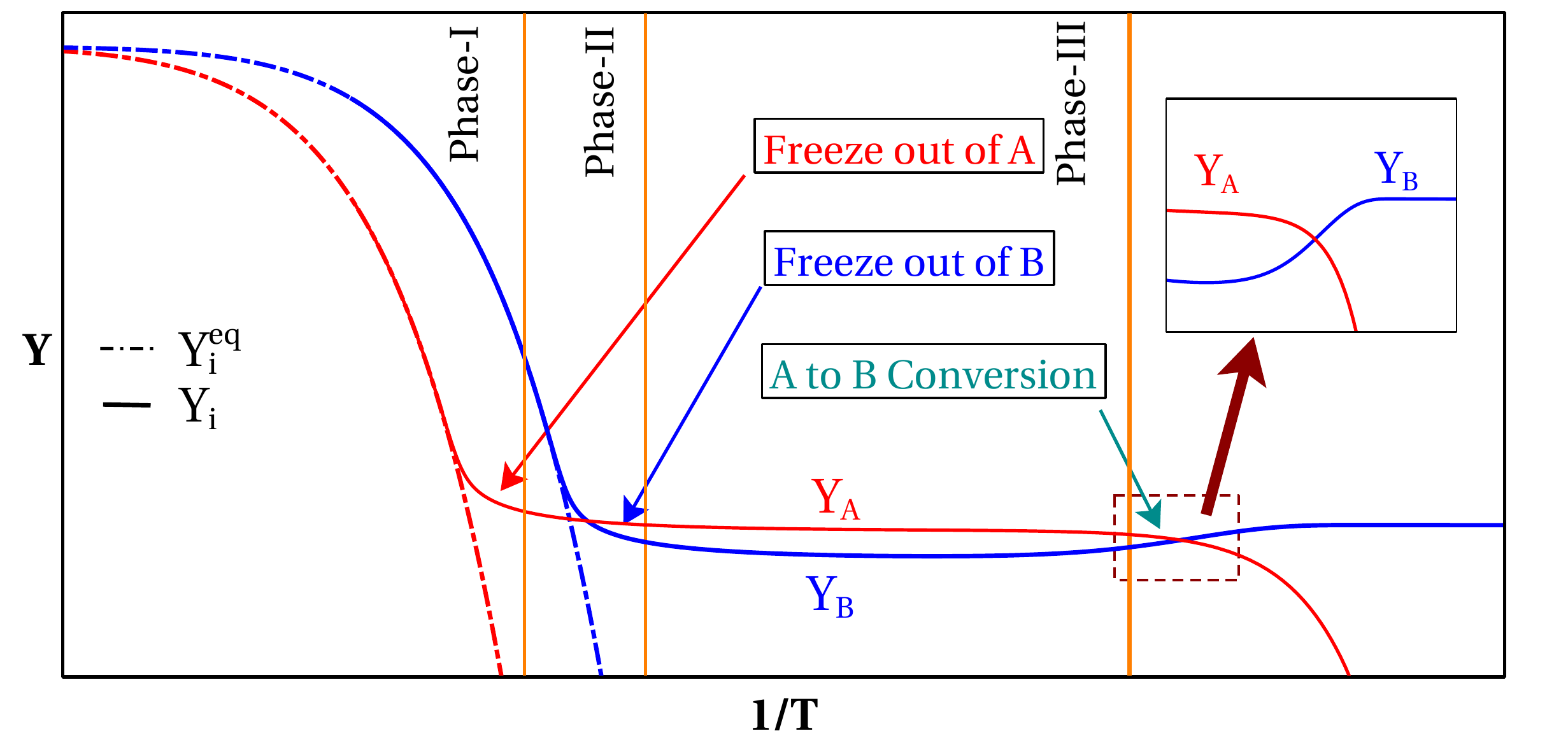}
    \caption{\it Cartoon diagram of the three phases of co-moving abundances evolving with $\frac{1}{T}$ for $M_A>M_B$. The solid and dashed line correspond to $Y_i$ and $Y_i^{eq}~(i=A,B)$ respectively.}
    \label{fig:cartoon_abund}
\end{figure}

\noindent The general framework of the DM production is shown by cartoon diagrams in Fig-\ref{fig:cartoon_int} and Fig-\ref{fig:cartoon_abund}.
Here {\bf $A$} is the heavier dark sector particle and $B$ plays the role 
of the stable DM candidate. 
In Fig-\ref{fig:cartoon_int} we show the interaction for such an oversimplified setup.
Initially, at very high temperatures both $A$ and $B$ were in thermal equilibrium with the bath particles with the help of the possible interactions $A~A \leftrightarrow {\rm SM~SM}$ and $B~B \leftrightarrow {\rm SM~SM} $ respectively.
First, the heavier dark sector particle $A$ is thermally produced which we call phase-I.
In the next step,  DM $B$  freezes out from the thermal bath which is named as phase-II. 
At this point, if there were no interaction between $A$ and $B$ both of them could serve as two component dark matter.
But here due to the presence of the non-thermal decay of $A$( $A\to B+{\rm SM}$) finally the number density of $A$ is further added up to the abundance of DM $B$ in  phase-III.
%and can fulfill the total observed DM abundance by PLANCK even below the TeV scale DM mass.
Note that in this case, particle A is not dominating the energy density of the universe at the time of decay. Hence we can neglect the entropy injection in the pre-BBN era due to the late decay of A \cite{Drees:2006vh}. In certain scenarios where the mother particle holds a significant portion of energy density, the decay products can have impact the freeze out abundances, as disscussed in ref.\cite{Giudice:2000ex} in contrast to our case.

For this general framework, we briefly discuss the numerical approach for evaluating the abundance of DM, $B$ ($M_A> M_B$). The evolution of co-moving density, $Y_i$ ($Y_i=n_i/s$ where $n_i,s$ are number density and the entropy density) for both the components $A$ and $B$ in the early universe
as a function of temperature can be described by the following coupled Boltzmann equation(BEQ), 
\begin{eqnarray}
G(z) \dfrac{dY_{A}}{dz} &=& -s \left< \sigma v \right>_{AA \leftrightarrow {\rm SM~SM}} \left(Y_{A}^2 -(Y_{A}^{eq})^2 \right) \nonumber\\
&&-\left< \Gamma_{A\rightarrow B~{\rm SM}} \right> \left(Y_{A}-\frac{Y_B}{Y_B^{eq}} Y_{A}^{eq}\right),\label{eq:A} \\
G(z) \dfrac{dY_{B}}{dz} &=& -s \left< \sigma v \right>_{BB \leftrightarrow {\rm SM~SM}} \left(Y_{B}^2 -(Y_{B}^{eq})^2 \right) \nonumber\\
&&+\left< \Gamma_{A\rightarrow B~{\rm SM}} \right> \left(Y_{A}-\frac{Y_B}{Y_B^{eq}} Y_{A}^{eq}\right).
\label{eq:B}
\end{eqnarray}
where the function $G(z)$ defined as: $G(z)=H(z)z  \left(1-\frac{1}{3} \frac{d \ln g_{s}(z)}{d\ln z}\right)^{-1}$. $z=M_{sc}/T$ is a dimensionless variable and $M_{sc}$ is an arbitrary mass scale to scale the temperature. $H$ is the Hubble expansion rate and is defined as 
$H=\sqrt{\frac{8\pi^3 g_{*\rho}}{90}}~ \frac{T^2}{M_{pl}}$ with $M_{pl}=\frac{1}{\sqrt{ G}}=1.22\times 10^{19}$ GeV \cite{Kolb:1990vq}.   
$g_{*\rho}$ and $g_{s}$ are the degrees of freedom associated with energy density and entropy density respectively\cite{Kolb:1990vq}. 
$ Y_{A}^{eq}~{\rm and}~ Y_{B}^{eq}$
are the equilibrium co-moving number density of the species $A$ and $B$ respectively.  
The thermal average cross-section of the annihilation process ${X~X \leftrightarrow {\rm SM~SM}}$ (with $X=A,B$) is denoted by $\left< \sigma v \right>_{X~X \leftrightarrow {\rm SM~SM}}$ and $\left< \Gamma_{A\rightarrow B~{\rm SM}} \right>$ signifies the thermally averaged decay of $A\to B+{\rm SM}$. The annihilation process $A~A\leftrightarrow {\rm SM~SM}$ is responsible to keep the particle $A$ in the thermal bath at the early universe while for $B$ it is $B~B\leftrightarrow {\rm SM~SM}$. 

%===================

\noindent The evolution of the co-moving number densities for such a general setup is shown in Fig-\ref{fig:cartoon_abund}. The blue and red lines indicate the number densities for species $A$ and $B$ respectively.
The dashed and solid lines correspond to $Y_i^{eq}$ and $Y_i~(i=A,B)$ respectively.
As the temperature drops below around $M_i$, respective $i$ particle density freezes out from the thermal bath and yields non-zero density, $Y_{i}$.
Due to the chosen parameter space, the heavier particle freezes out early 
followed  by the lightest dark sector particle which is named as phase-I and phase-II respectively in Fig-\ref{fig:cartoon_abund}. 
And finally due to the chosen dark Yukawa coupling the density of heavier species $A$ dilutes to $B$ which is phrased as phase-III in the above Fig-\ref{fig:cartoon_abund}. Therefore one can take the final abundance of B approximately as $Y_B^{z\to\infty} \simeq Y_A(z^{\rm FO}_A)+Y_B(z^{\rm FO}_B)$ where $Y_i(z^{FO}_i)$ ($i=A,B$) be the co-moving number density of $i^{\rm th}$ particle at the time of freeze-out $z^{FO}_i=\frac{M_{\rm sc}}{T_i^{\rm FO}}$. So the relic density of $B$ will be
\bea
\Omega_{B} h^2=2.755 \times 10^{8} \times \left(\frac{M_B}{\rm GeV}\right) Y_B^{z\to\infty}.
\eea
Note that the DM observed abundance measured by the Planck collaboration to be $\Omega_{\rm DM} h^2=0.120\pm0.001$ \cite{Planck:2018vyg}.
%==============================================
\section{Dirac Dark Matter}
\label{sec:diracdm}
%===============================================
To realize the hybrid setup of DM production discussed in the earlier section we extend the SM particle content with a vector like  $SU(2)_L$ lepton doublet, $\Psi=(\psi^0~~\psi^-)^T$ 
and an additional real scalar singlet, $S$ ,where the neutral component $\psi^0$ can act as a DM.
For the minimal case (without $S$), the lepton doublet is the only dark sector field, and the neutral component serves as a stable DM candidate, stabilized by an extended $\mathcal{Z}_2$ symmetry.
It  has been found that the neutral component of lepton doublet alone can not accommodate the observed relic for DM mass $M_{\Psi}\lesssim 1.2$ TeV \cite{Bhattacharya:2018fus}. 
The reason behind the under-abundance issue is the large interaction rate with SM particles due to the gauge interactions.
For $M_{\Psi}\gtrsim 1.2$ TeV DM $\psi^0$ becomes over-abundant.
To probe the under-abundant lepton doublet DM mass(sub-TeV) region we propose the hybrid setup and augment the particle content with additional dark sector particle $S$.
Both the dark states($\Psi$ and $S$) are odd and the SM particles are even under the additional $\mathcal{Z}_2$ symmetry to ensure the stability of DM.
The relevant part of the Lagrangian involving DM along with heavier dark scalar $S$ is described as follows,
%\begin{widetext}
\bea
\mathcal{L} &\supset& \underbrace{\overline{\Psi}\Big[i \gamma^\mu \Big( \partial_\mu - i g_2 \frac{\sigma^a}{2} W_{\mu}^a-i g_1 \frac{Y_{\psi}}{2} B_{\mu}\Big)- M_\Psi\Big] \Psi}_{\mathcal{L}_\Psi}~~\nonumber \\
&& \underbrace{-y_s \sum_{i=e,\mu,\tau} \overline{L_i}~P_R\Psi S +h.c.}_{\mathcal{L}_{\Psi-S}} \nonumber \\
&& 
\underbrace{-\frac{{M_S}^2}{2} S^2 -\frac{\lambda_{S \Phi}}{2} S^2 \Big(\Phi^\dagger\Phi -\frac{v^2}{2} \Big)-\frac{\lambda_S}{4!}S^4 }_{\mathcal{L}_S} 
\label{eq:mastlag}
\eea
%\end{widetext}
%
The first term of the Lagrangian, $\mathcal{L}_\Psi$ involves the interaction between the $SU(2)_L$ lepton doublet and the SM gauge bosons. Here $g_2$ and $g_1$ are the gauge coupling corresponding to the SM $SU(2)_L$ and $U(1)_Y$ gauge groups respectively while the corresponding gauge bosons are denoted by $W_\mu$ and $B_\mu$. 
$L_i(i=e,\mu,\tau)$ represents SM lepton doublet whereas $\Phi$ is SM Higgs doublet.   
The gauge interactions are responsible for the thermal freeze out of the neutral component lepton doublet DM, $\psi^0$. $M_\Psi$ is the bare mass of the lepton doublet.  
The masses of the neutral($\psi^0$) and charged($\psi^\pm$) dark leptons can be expressed as\cite{Thomas:1998wy},
\begin{equation}
    M_{\psi^0}=M_\Psi,~~~M_{\psi^\pm}=M_\Psi+ \delta m .
    \label{eq:qc}
\end{equation}
 
Here $\delta m$ is mass the splitting between $\psi^\pm$ and  $\psi$, generated from 1 loop quantum correction as\cite{Thomas:1998wy}:

\begin{equation}
\delta m =\frac{\alpha}{2} M_Z f\Big(\frac{{M_\Psi}^2}{M_Z^2}\Big) 
\nonumber
\end{equation}

\begin{equation}
{\rm with}~f(r)=\frac{r}{\pi}\int^1_0 dx (2-x) ln\Bigg[ 1+ \frac{x}{r\sqrt{(1-x)^2}}\Bigg],
\end{equation}
and generally is of the order of pion mass ($\delta m \sim m_{\pi}$). 
\noindent $\mathcal{L}_S$ is the interaction Lagrangian for the dark real scalar singlet,$S$ where $S$ is interacting with
SM through Higgs portal interactions as $S^2 \Phi^\dagger \Phi$. 
With $M_S >0$ and $\lambda_S >0$, the scalar field $S$ does not acquire any vacuum expectation value (vev), thus the $\mathcal{Z}_2$ symmetry remains intact. $M_S$ is the physical mass of $S$. 

\noindent
The last term of the Lagrangian, $\mathcal{L}_{\Psi-S}$ involving the Yukawa interaction between dark sector particles($\Psi$ and $S$) and the SM leptons is the most important term for our analysis. 
Depending on the value of $y_s$ the Yukawa interaction $\mathcal{L}_{\Psi-S}$ will give rise to late time re-population for the lightest dark sector particle ($\psi^0$ or $S$). 
We assume the mass of the scalar $S$ is heavier than the masses of the fermion states i.e. $M_S > M_{\psi^\pm}+M_\ell$. With the mass hierarchy the heavier dark sector particles can decay into the lightest state, $\psi^0$ as $S \to \psi^0~\overline{\nu}$ and $S\to \psi^-~\ell^+ \to \psi^0~\pi^- \ell^+$, where $l$ denotes charged leptons. As a result the lightest neutral fermion, $\psi^0$ acts as a stable DM candidate thanks to the Yukawa coupling, $y_s$.
We choose small values of $y_s$ such that the decays start after both $S$ and $\psi^0$ freeze out from the thermal bath and also before the BBN which will be addressed shortly.
Mapping the dark sector with the discussion made in sec-\ref{sec:gen}
one notices that $S$ and $\psi^0$ resemble $A$ and $B$ respectively.
For simplicity, we consider a universal Yukawa coupling of $S$ ($y_s$)  with the lepton sector.
The corresponding decay widths of the heavy dark scalar,$S$ 
to $\psi^0 + \nu$ and $\psi^\pm \ell^\mp $ is given by the following:
\bea
\Gamma_{S\to \psi^0 \overline{\nu}}&=& y_s^2 \sum_{\ell=e,\mu,\tau} \frac{1}{16 \pi M_S^3}
~(M_S^2-M_{\psi^0}^2)^{2} \nonumber \\
\Gamma_{S\to \psi^{-} {\ell}^+}&=& y_s^2 \sum_{\ell=e,\mu,\tau} \frac{1}{16 \pi M_S^3}~\{M_S^2-(M_{\psi^\pm}-M_{\ell})^2\} \nonumber \\
&& \times \sqrt{\{M_S^2-(M_{\psi^\pm}+M_{\ell})^2\}}
\nonumber \\ && \times \sqrt{\{M_S^2-(M_{\psi^\pm}-M_{\ell})^2\}}~.
\label{eq:dcy}
\eea

We shall now identify the relevant parameters which control  the phenomenology of lepton doublet DM in this framework. The model has mainly four relevant free parameters and they are as follows:
\bea
\{M_S,~M_\Psi,~\lambda_{S\Phi},~y_s\}. 
\eea
In the rest of our discussions, we consider that 
$M_S > M_{\Psi}$, 
with $\psi^0$ serving as a DM. In addition to the aforementioned free 
parameters, the interaction between $\Psi$ and the Standard Model (SM) sector 
is controlled by the SM gauge couplings ($g_2$ and $g_1$). The interaction 
between $S$ and the SM is determined by the Higgs portal coupling, 
$\lambda_{S\Phi}$. The Yukawa coupling $y_s$ plays a significant role in 
determining the repopulation of dark matter $\psi^0$ from $S$, as 
well as setting the lifetime of $S$ ($\tau_S$).

Before delving into the detailed analysis of dark matter (DM), 
let us first provide a brief overview of various experimental limits 
imposed on the model parameters. For $M_{\Psi} < M_Z/2$, the invisible 
decay width of the $Z$ boson receives additional contribution beyond what
is predicted by the SM. However, the observational data on the invisible 
decay width of the $Z$ agrees remarkably well with the SM prediction, 
thus strongly necessitating $M_{\Psi} > M_Z/2 $~\cite{ParticleDataGroup:2020ssz}. Additionally, 
LEP-2 already excluded masses for exotic charged 
fermions, $M_{\psi^\pm}$, below $\sim$ 102.7 GeV \cite{DELPHI:2003uqw}. In this framework, 
both the charged and neutral leptons are degenerate, 
meaning $M_{\psi^\pm}\approx M_{\psi^0} \equiv M_{\Psi}$. As a result, the contribution of the lepton doublet to the EW precision parameters(S, T, and U) is consistent with the observed bound\cite{Cynolter:2008ea}.
Furthermore, the Higgs invisible decay width, as measured by LHC 
\cite{CMS:2018yfx}, imposes constraints on the Higgs portal coupling 
$\lambda_{S\Phi}$ when $M_S< m_h/2 $. However, in this framework we 
consider that $M_S > M_{\Psi}$, thus rendering the Higgs invisible decay 
constraint inapplicable. Hence, in our discussion, we establish a 
lower bound on the mass parameters as follows: $M_S > M_{\Psi} > 102.7$ GeV.
We will also consider the limits imposed by the observed 
DM relic abundance $\Omega_{\rm DM}h^2=0.120\pm0.001$ by the PLANCK 
Collaboration~\cite{Planck:2018vyg}, as well as direct search constraints 
from XENON \cite{XENON:2017vdw}, PANDA \cite{PandaX-II:2016vec}, and recent 
LZ data \cite{LZ:2022ufs}, along with indirect search constraints from 
FERMI-LAT \cite{Fermi-LAT:2016uux} and MAGIC \cite{MAGIC:2016xys}, on the model 
parameters.

%================================
%==================================
The neutral component of the lepton doublet, $\Psi$ with the mass hierarchy $M_\Psi < M_S$ behave as a stable DM candidate. 
The relevant Boltzmann equations are
%\begin{widetext}
\begin{eqnarray}
G(x) ~\dfrac{dY_{S}}{dx} &=& -{s} ~\left< \sigma v \right>_{S} \left(Y_{S}^2 -(Y_{S}^{eq})^2 \right) \nonumber\\
&&-{\left< \Gamma_{S\rightarrow \Psi}  \right>}  \left(Y_{S}-\frac{Y_{\Psi}}{Y_{\Psi}^{eq}} Y_{S}^{eq}\right),\label{eq:S}\\
G(x) ~\dfrac{dY_{\rm \Psi}}{dx} &=& -{s}\left< \sigma v \right>_{\rm \Psi}^{\rm eff} \left(Y_{\Psi}^2 -(Y_{\Psi}^{eq})^2 \right) \nonumber\\
&&+{\left< \Gamma_{S\rightarrow \Psi}\right>}  \left(Y_{S}-\frac{Y_{\Psi}}{Y_{\Psi}^{eq}} Y_{S}^{eq}\right) ,
\label{eq:cbeq}
\end{eqnarray}
%\end{widetext}
where $x=M_{S}/T$ is the dimensionless variable.
$ Y_{S}$ is the comoving abundances of $S$.
Note that in eq.\eqref{eq:cbeq} we write the Boltzmann equation for $Y_\Psi$ signifying the total abundances of $\psi^0$ and $\psi^\pm$. 
But due to the presence of strong co-annihilation between nearly mass degenerate charged and 
a neutral component of $\Psi$ like $\psi^0 \psi^\pm, \psi^+ \psi^- \to {\rm SM~SM}$, $Y_{\psi^\pm}$ becomes zero at the moment of freeze out leaving only non zero $Y_{\psi^0}$ i.e. $Y_\Psi^{x_F}=Y_{\psi^0}^{x_F}+Y_{\psi^\pm}^{x_F}\simeq Y_{\psi^0}^{x_F}$.
%, denotes the co-moving number density of $\psi$.  
Here, $x_F=M_S/T_{\rm FO}$ and $T_{\rm FO}$ denotes the freeze out temperature of respective species. 
$Y_{\Psi}^{eq}$ and $ Y_{S}^{eq}$ are the equilibrium co-moving number density of $\Psi$ and $S$ respectively. 
 $\left< \sigma v\right>_S$ denotes the thermal average cross-section of $S$ to the bath particles(SM), $S~S\to {\rm SM~SM}$. 
 And $\left< \sigma v\right>_\Psi^{\rm{eff}}$ represents the effective thermal average cross-section of the dark lepton doublet  associated 
 with the annihilation processes, $\psi^0~\psi^0 \to {\rm SM~SM}$ and the co-annihilation processes, $\psi^0 \psi^\pm, \psi^+ \psi^- \to {\rm SM~SM}$. 
 The effective thermal average cross-section, $\left< \sigma v\right>_\Psi^{\rm{eff}}$ can be expressed as follows \cite{Griest:1990kh,Edsjo:1997bg}:
\begin{widetext}
\bea
\left< \sigma v\right>_\Psi^{\rm{eff}}&=& \frac{g_{\psi^0}^2}{g_{\rm eff}^2} \left< \sigma v \right>_{\psi^0\psi^0} +\frac{g_{\psi^\pm}^2}{g_{\rm eff}^2} \left< \sigma v \right>_{\psi^+ \psi^-}(1+\delta_{\psi^\pm})^3e^{-2\zeta\delta_{\psi^\pm}}
+\frac{2 g_{\psi^0} g_{\psi^\pm}}{g_{\rm eff}^2} \left< \sigma v \right>_{\psi^0\psi^\pm}(1+\delta_{\psi^\pm})^{\frac{3}{2}}~ e^{-\zeta~\delta_{\psi^\pm}} ,
\eea
\bea
{\rm with}~~ g_{\rm eff}=g_{\psi^0}+g_{\psi^\pm}(1+\delta_{\psi^\pm})~ e^{-\zeta\delta_{\psi^\pm}},~~\zeta=M_{\psi^0}/T~~{\rm and}~~ \delta_{\psi^\pm}=\frac{M_{\psi^\pm}-M_{\psi^0}}{M_{\psi^0}} \nonumber.
\eea
\end{widetext}
The internal degrees of freedom  $g_{\psi^0}$ and $g_{\psi^\pm}$ are associated with the  dark lepton states, $\psi^0$ and $\psi^\pm$ respectively. The thermal average of the total decay width of $S$, ($\Gamma(S\to\ell^\pm \psi^\mp) +\Gamma(S\to \psi^0+\bar{\nu})$) is denoted by $\left< \Gamma_{S \to \Psi } \right>$. It is important to note that for numerical analysis we have adopted the $\left< \sigma v\right>_\Psi^{\rm{eff}}$ as a function of temperature for a given set of parameters using the open-code {\bf micrOmega}\cite{Belanger:2014vza}. To generate the model files for {\bf micrOmega}, we first implement the model in the public code {\bf FeynRule}\cite{Alloul:2013bka}. Using the $\left< \sigma v\right>_\Psi^{\rm{eff}}(T)$, the analytical expression of $\left< \sigma v \right>_S$ for the annihilation processes of the Higgs portal $S$, $S~S\to {\rm SM~SM}$\cite{Bhattacharya:2016ysw}, and the analytical expression of $\left< \Gamma_{S \to \Psi } \right>$(in eq.\eqref{eq:dcy}), we solve the above coupled BEQ as described in eq.\eqref{eq:S} and eq.\eqref{eq:cbeq}.  

The evolution of the co-moving number densities of $S$ and $\psi^0$ are described by eq.\eqref{eq:S} and  
eq.\eqref{eq:cbeq} respectively. 
The first term on the right-hand side (R.H.S)
of eq.\eqref{eq:S} decides the freeze out of  $S$ as the temperature drops $T<M_S$. 
Similarly, the first term on the R.H.S of eq.\eqref{eq:cbeq} decides the freeze out of  $\psi^0$ as the temperature drops $T<M_\psi^0$ with $Y_{\psi^\pm}=0$ at the time of freeze out.
However, the presence of a tiny Yukawa coupling $y_s$ among fermion DM, heavy dark scalar, and SM leptons gives rise to the late time 
decay of  $S$ into fermion DM and eventually dilutes the number density of $S$ as depicted by the second term of eq.\eqref{eq:S}. 
Similar to this, the  second term in the R.H.S of the eq.\eqref{eq:cbeq} governs the re-population of $\psi^0$ from the decay of $S$.
Note that from the late decay of $S$, $\psi^\pm$ is also produced. 
But due to the small mass splitting between $\psi^\pm$ and $\psi^0$ ($\delta m\sim \mathcal{O}(m_\pi)$ ), $\psi^\pm$ promptly decays to 
$\psi^0$ ($\psi^\pm \to \psi^0+\pi^\pm$)  \cite{Cirelli:2005uq} and eventually all the number densities get converted to $\psi^0$ density. 
As previously stated, we set the coupling $y_s$ and the mass $M_S$ in such a way that the decay is active after the freeze out of $\psi^0$ and after the decay gets completed $Y_S$ is totally converted to $Y_{\psi^0}$.
It is worth noting that in the absence of the Yukawa interaction, both $\psi^0$ and $S$ becomes stable and can act as two component DM\cite{Bhattacharya:2018cgx}. However, we are interested in sub-TeV fermion doublet DM, which can accommodate the entire observed DM abundance by PLANCK\cite{Planck:2018vyg} in the presence of a heavy dark scalar, $S$, with the tiny Yukawa coupling $y_s$.
Before going to numerical solutions of $Y_S, Y_{\psi^0}$ we mention some important issues regarding $y_s$ in the next paragraph.

%=============================
\begin{figure}[tbh!]
    \centering
    \includegraphics[scale=0.38]{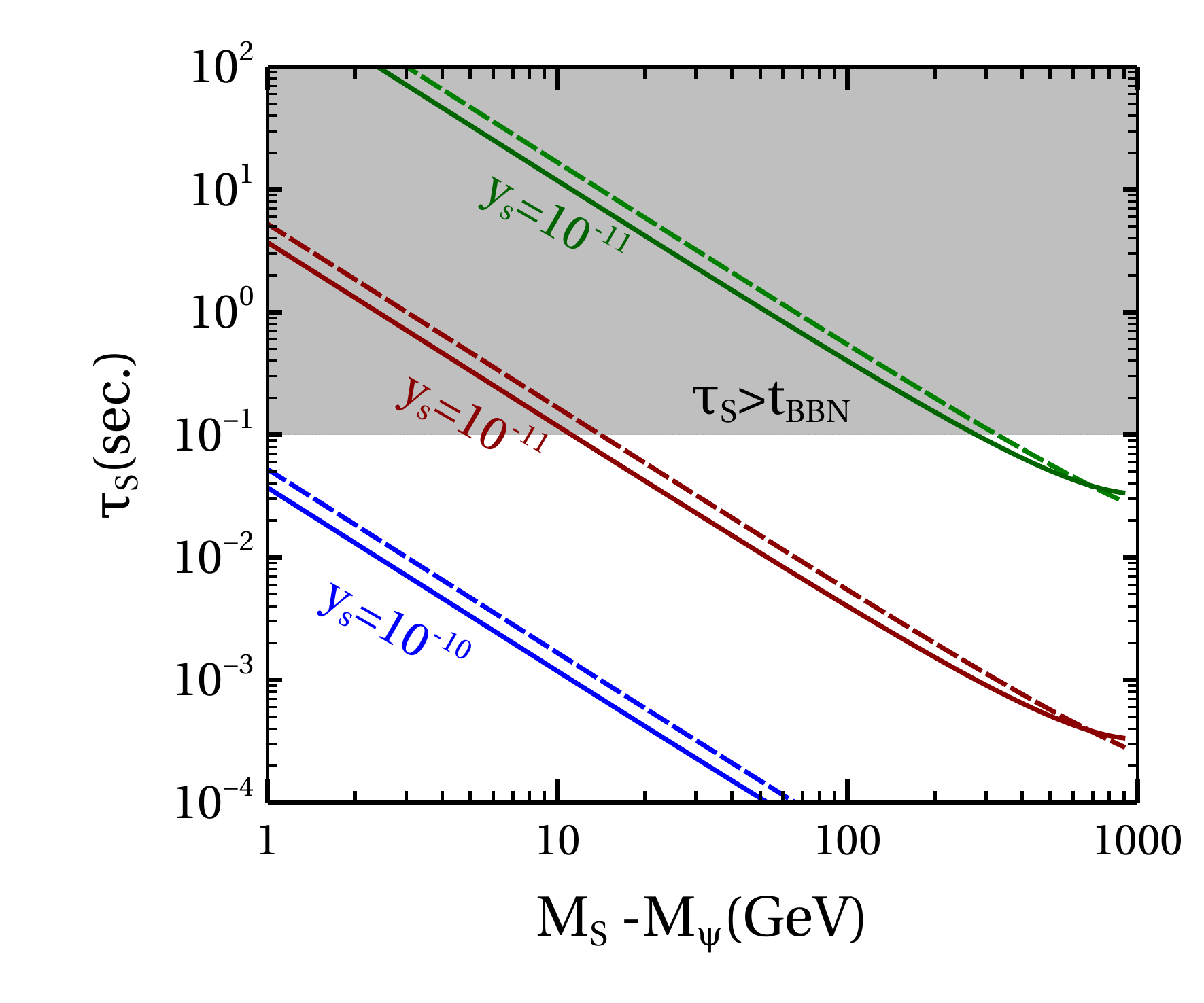}
    \caption{\it Variation of $\tau_S$ with $M_S-M_\Psi$ for different values of $y_s=10^{-10},10^{-11},10^{-12}$ shown in blue,red and green lines respectively. The solid and dashed line correspond to $M_S=1$ TeV and $2$ TeV respectively. The shaded region corresponds to where the decay will be active after BBN.}
    \label{fig:lyftm}
\end{figure}

%==============================
While dealing with such late time decays, $S\to \ell^\pm \psi^\mp,~\overline{\nu}~\psi^0(\nu~\overline{\psi^0})$, one should be careful so that the decay products do not alter the predictions of BBN. 
The charged fermion $\psi^{\pm}$  decays further to $\pi^{\pm}$ and $\psi^0$ with $97\%$ branching ratio and even the pions can decay to leptons.
These excess pions and leptons may interact with nucleons during BBN and tamper the abundances of light elements attracting strong constraints \cite{Feng:2003uy}.
To be on the safer side and simplify our analysis, we restrict our analysis by demanding that the decay is completed at a much higher temperature than BBN (i.e. %$T_{S\to \psi^0} \gg T_{\rm BBN}$ or 
$\tau_{S} \ll t_{\rm BBN}$). 
In Fig-\ref{fig:lyftm}, we show the parameter space in $M_S-M_{\Psi}$ vs. $\tau_S$ plane where $\tau_S$ is the life time of $S$.
Here, we consider three different values of $y_s=10^{-10},~10^{-11},~10^{-12}$
shown by blue, red, and green lines respectively.
The solid and dashed line correspond to $M_S=1000$ GeV and $2000$ GeV
respectively.
The gray shaded region corresponds to the parameter space where the decay of $S$ will occur after BBN.
According to eq.\eqref{eq:dcy}, the lifetime of $S$ is inversely proportional to $y_s$ and $M_S-M_\Psi$.
This feature is evident from the aforementioned figure.
%Therefore, for smaller $\Delta M_S$ one has to choose higher $y_s$ to meet the aforementioned criteria for life time.
We also want the coupling $y_s$ to be such that the decay of $S$ starts non-thermally after the freeze out of $\psi^0$ and for that reason, $y_s$ should be $\lesssim 10^{-9}$.
With $y_s=10^{-10}$ the decay of $S$ also gets completed before BBN. 
However, apart from setting the lifetime, $y_s$ plays no role in deciding the relic which we will discuss in the following paragraph \cite{Coy:2021sse}. 
%==========================
\begin{figure}[tbh]
    \centering
    \subfigure[\label{a}]{
    \includegraphics[scale=0.38]{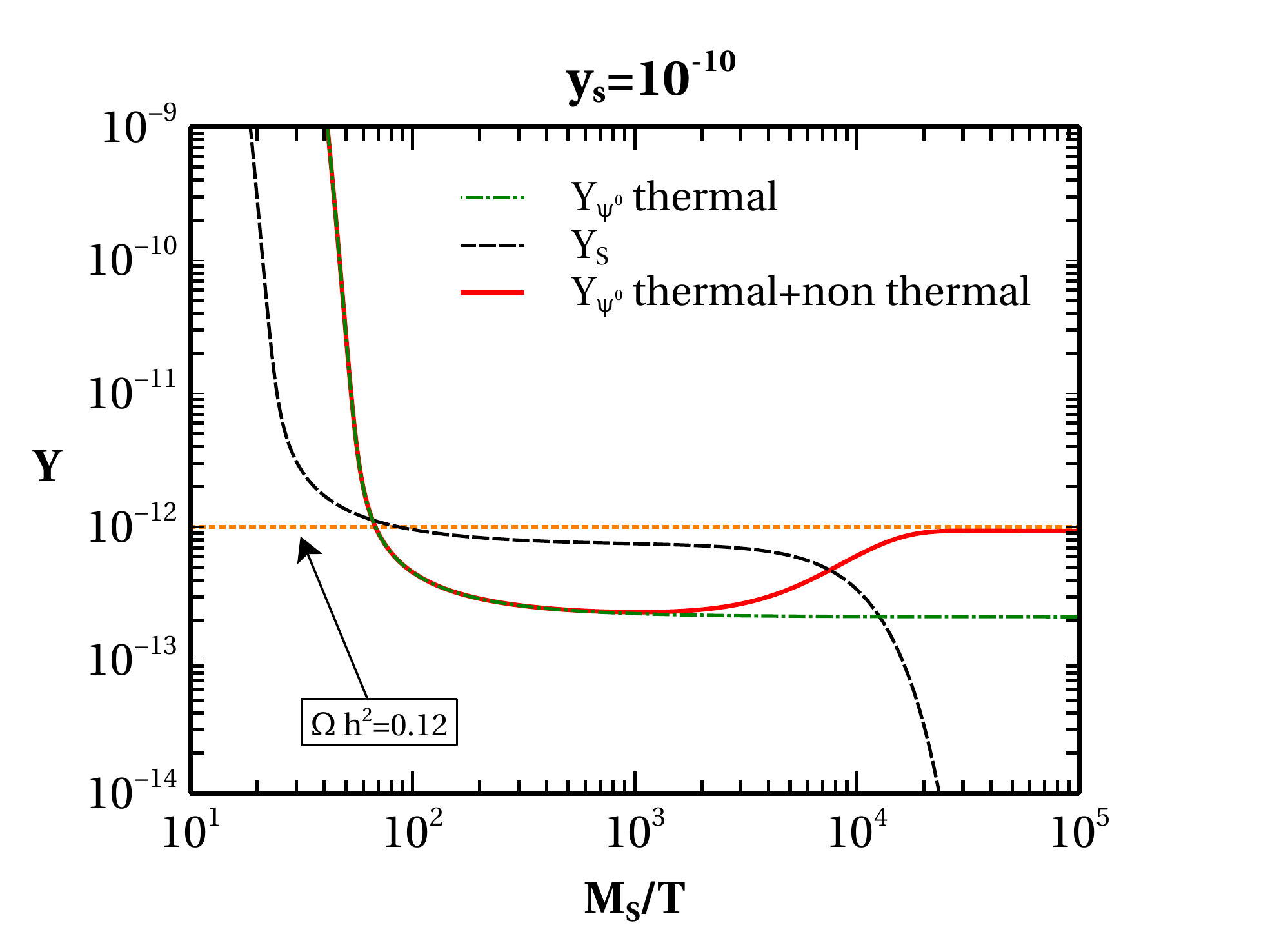}}
    \subfigure[\label{b}]{
    \includegraphics[scale=0.38]{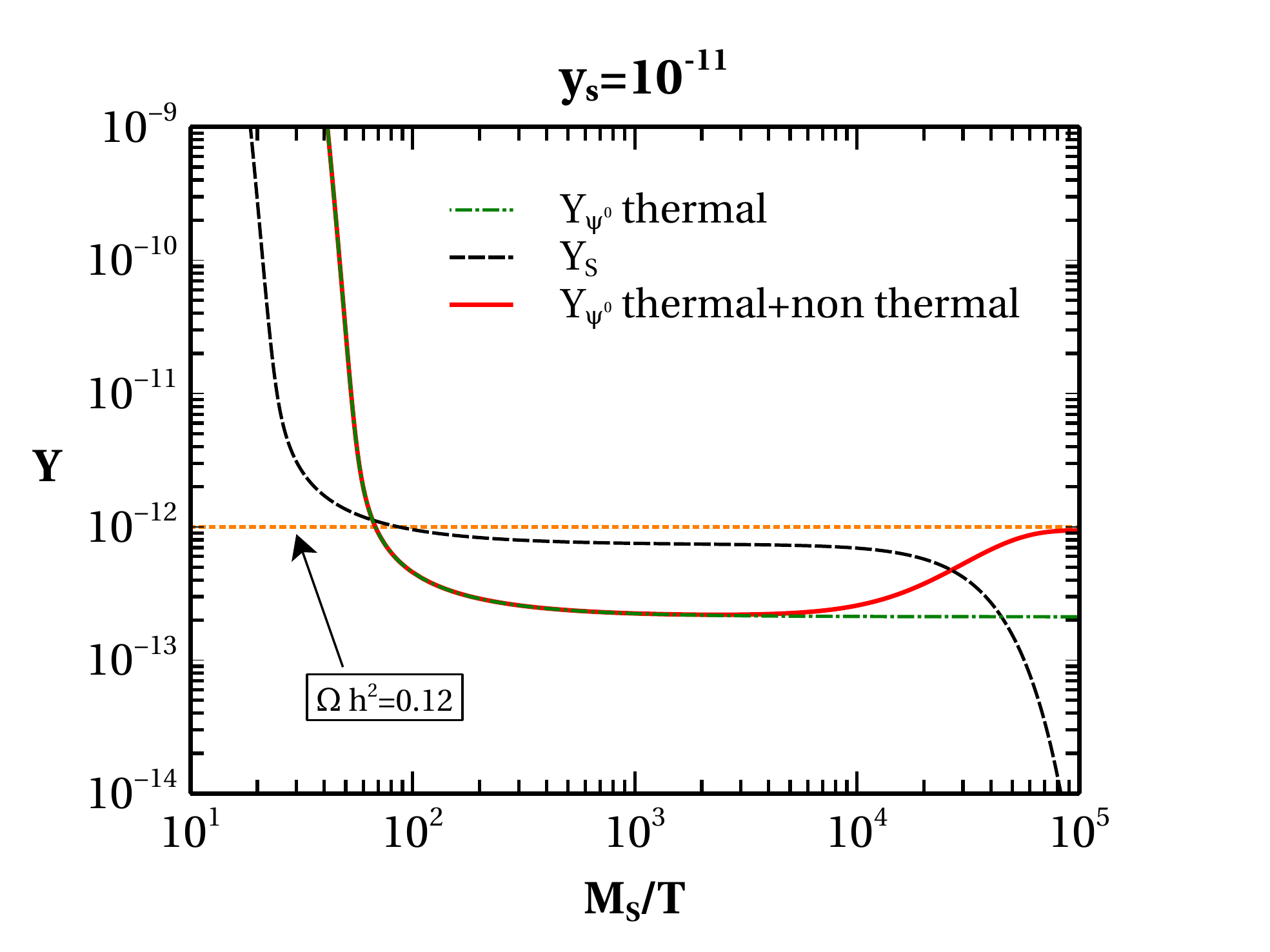}}
    \caption{\it Comoving abundances as a function of inverse temperature for $M_{S}=1000$ GeV, $M_{\Psi}=400$ GeV and $\lambda_{S \Phi}=0.245$. We consider $y_s=10^{-10}$ in (a) and $y_s=10^{-11}$ in (b) for demonstration. The black dashed and red solid lines signify $Y_S$ and $Y_{\psi^0}$ respectively. The green dashed dot line correspond to $Y_{\psi^0}$ in absence of the dark Yukawa coupling i.e. $y_s=0$. The orange dotted line corresponds to comoving abundances leading to observed relic density for $M_{\Psi}=400$ GeV.}
    \label{fig:dm1}
\end{figure}

We shall now discuss how the model parameters affect the DM number density obtained from the BEQs in eq.\eqref{eq:S} and eq.\eqref{eq:cbeq}.
In Fig-\ref{fig:dm1}, we showcase the evolution of comoving abundances of $S$ and $\psi^0$ with the dimension less quantity $x=M_S/T$. 
For illustration, we choose $M_S=1000$ GeV, $M_{\psi^0} \simeq M_{\Psi}=400$ GeV and $\lambda_{S \Phi}=0.245$.
We take $y_s=10^{-10}$ and $10^{-11}$ for Fig-\ref{a} and \ref{b} respectively, to be compatible with the cosmological limitation described above.
The comoving abundances of $S$ ($Y_S$) and $\psi^0$ ($Y_{\psi^0}$) are denoted by the black dashed and the red solid lines.
The orange dotted line corresponds to comoving density leading to observed relic density by PLANCK\cite{Planck:2018vyg}.
The green dashed line corresponds to $Y_{\psi^0}$ in the absence of late-time DM production via non-thermal decay (i.e. $y_s=0$). 
We can see from the figures that both $S$ and $\psi^0$ freeze out from the thermal bath, and after some time, $S$ starts to decay, resulting in additional number densities of $\psi^0$. As a result, we see an increase in $Y_{\psi^0}$ leading to satisfying the observed relic, whereas in the absence of late DM production (green dashed line), it fails to fulfill the observed relic. 
Comparing Fig-\ref{a} and \ref{b}, we notice that with decreasing $y_s$ the late DM production takes place at a comparably later time (lower $T$). However, $y_s$ plays no role in deciding the abundance of DM other than determining the lifetime of the $S$, as  $S$ ultimately decays to DM $\psi^0$. 
%However, apart from setting the life time of $S$, $y_s$ has no role in deciding the abundances as ultimately all the $S$ particle decays to $\psi^0$ and $\psi^{\pm}$.Eventually all the $\psi^{\pm}$ also decays to $\psi^0$ upto a good approximation. 
So, at the later time, we can just assume $Y^{x\to \infty}_{\psi^0} \equiv  Y_{S}(x_{S}^{\rm FO})+ Y_{\Psi}(x_{\Psi}^{\rm FO})$.
To avoid entering in the vicinity of BBN, we take a fixed $y_s=10^{-10}$ throughout the analysis. Therefore,
with this kind of setup, one can easily accommodate sub-TeV fermion doublet DM with the appropriate choice of $M_S$ and $\lambda_{S \Phi}$.

%==========================
\begin{figure}[tbh]
    \centering
     \subfigure[\label{ax}]{
    \includegraphics[scale=0.35]{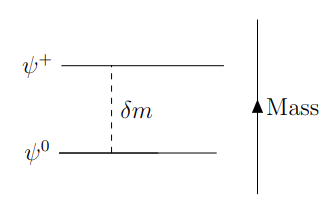}}
    \subfigure[\label{bx}]{
        \includegraphics[scale=0.32]{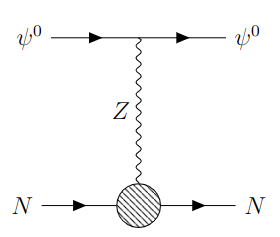}}
    \caption{\it (a) Mass spectrum of the Dirac lepton doublet $\Psi$. (b)Feynman diagram for $Z$ mediated elastic scattering between Dirac DM $\psi^0$ and nucleon (proton/neutron) $N=\{p,n\}$.}
    \label{fig:mx}
\end{figure}

However, despite the fact that the DM $\psi^0$ can easily  fulfill the total observed relic density within the sub-TeV mass range with such minimal setup, the gauge-mediated direct search imposes a strong constraint on DM mass \cite{Bhattacharya:2018fus}.
The mass spectrum of $\Psi$ and the relevant interaction of direct detection are shown in Fig-\ref{ax} and Fig-\ref{bx} respectively. 
%The reason behind this is the strong $Z$ mediated gauge interaction with nucleons. 
The current bound from direct detection experiments by 
LUX \cite{LUX:2016ggv}, XENON-1T \cite{XENON:2017vdw} completely rule out the possibility of Dirac DM $\psi^0$ even upto $2$  TeV. 
Nevertheless, such a scenario with a dark doublet can be  revived  with pseudo-Dirac DM with the help of an additional scalar triplet as we will discuss in the next section.

%==============================================
\section{Pseudo-Dirac Dark Matter}
\label{sec:psudodirac}
%===========================================
The  elastic direct detection cross-section bound for DM($\psi^0$) can be easily evaded if the DM turns out to be a pseudo-Dirac state, in which case the $Z$ mediated neutral current vanishes. Such type of scenario can be realized by exploiting the pseudo-Dirac nature of DM where the state $\psi^0$ splits into two Majorana states. With such motivation, we introduce an additional 
$SU(2)_L$ scalar triplet $\Delta$ (with hypercharge, $Y=2$) that creates the pseudo-Dirac mass splitting. 
The relevant parts of the Lagrangian involving $\Delta$ are given by,   
\bea
\mathcal{L} &\supset& \mathcal{L}_{\Delta+{\rm SM}}+\mathcal{L}_{S\Delta} -\frac{y_{_\psi}}{\sqrt{2}} \bar{\Psi^c}i \sigma^2\Delta\Psi + h.c.  ~~.
\label{eq:psi1}
\eea

\noindent The Yukawa interaction between the vector-like lepton doublet ($\Psi$) and the additional scalar triplet($\Delta$) is proportional to $y_\psi$ and plays a crucial role in generating pseudo-dirac splitting of $\psi^0$. However, there are several motivation for introducing a $SU(2)_L$ scalar triplet in SM, commonly known as the Type-II seesaw model in literature \cite{Cheng:1980qt,Arhrib:2011uy,Mohapatra:1980yp,Schechter:1980gr}. One of the main motivations for this is to address the explanation of neutrino masses, which have been studied extensively so far\cite{Cheng:1980qt,Mohapatra:1980yp}. In this paper, we are only interested in the dark matter analysis. We briefly describe the relevant part of the interaction Lagrangian in this context, $\mathcal{L}_{\Delta+{\rm SM}}$ in the appendix-\ref{sec:apxDelta}. The interaction between the dark scalar $S$ and $\Delta$ in the modified scenario is described by the Lagrangian 
\bea
\mathcal{L}_{S\Delta}=\mathcal{L}_S-\frac{\lambda_{S\Delta}}{2} S^2 \left({\rm Tr}[\Delta^\dagger \Delta]-\frac{{v_\Delta}^2}{2}\right),
\eea
where the interaction $\mathcal{L}_S$ is defined earlier in eq.\eqref{eq:mastlag} with $v$ replaced by $v_d$ satisfying $\sqrt{v_d^2+2 {v_\Delta}^2}=246$ GeV. To avoid additional contribution in the thermal abundance of $S$ from $\Delta$, we assume $\lambda_{S\Delta}$ to be zero and $M_S < M_{\Delta}/2$ which simplifies our analysis. 

\noindent The scalar triplet $\Delta$ does not acquire any vacuum expectation value(vev). Although electroweak symmetry breaking (EWSB) causes the formation of an induced vev $v_\Delta$ along the neutral CP even field direction, which can modify the EW parameters. 
The precision measurements of the EW observable constrain the $\rho$ parameters $\rho=1.00038 \pm 0.00020$\cite{ParticleDataGroup:2020ssz}, putting an upper bound on $v_\Delta$ as $v_\Delta \lesssim 2.6$ GeV at the $3\sigma$ level. There also exists a lower bound on  $v_\Delta \gtrsim 10^{-9}$ GeV from lepton flavour violation \cite{MEG:2016leq}.

\noindent
 Here, the $v_\Delta$ induces a  tiny Majorana mass, $y_{\psi} v_{\Delta}$ to the Dirac state $\psi^0$ thanks to the Yukawa interaction mentioned eq.\eqref{eq:psi1}. 
 This Majorana mass term leads to the mass splitting of the $\psi^0$ state into two physical states, $\psi_1$ and $\psi_2$, after proper diagonalization, as discussed in equation  eq.\eqref{eq:pstate}
\footnote{Equivalently one can write dimension-5 operator like ${1}/{\Lambda} ~\overline{\Psi} \tilde{H} \tilde{H}^T \Psi^c $ in order to generate the pseudo-Dirac mass splitting. For detail analysis see ref.\cite{Essig:2007az}} . 
 The physical states and their corresponding masses for the lepton doublet are described below:
\bea
\psi_1&=&\frac{i}{\sqrt2}({\psi^0}^c-\psi^0)~{\rm with}~M_{\psi_1}= \Big( M_\Psi -  v_{_\Delta} y_{_\psi} \Big),\nonumber \\
\psi_2&=&\frac{1}{\sqrt2}({\psi^0}^c+\psi^0)~{\rm with}~M_{\psi_2}= \Big( M_\Psi +  v_{_\Delta} y_{_\psi} \Big), \nonumber \\
{\rm and} ~\psi^\pm && ~~~~~~~~~~~~~~~~~~~~{\rm with}~M_{\psi^\pm}=M_{\Psi}+\delta m~.
\label{eq:pstate}
\eea

\begin{figure}[tbh]
    \centering
    \subfigure[\label{c}]{
    \includegraphics[scale=0.34]{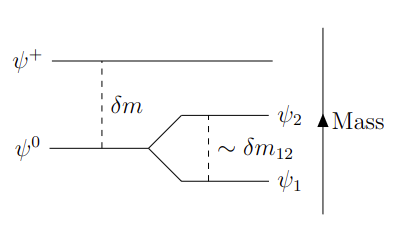}}
    \subfigure[\label{d}]{
    \includegraphics[scale=0.32]{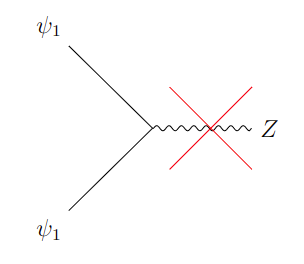}}
    \caption{\it (a)Mass spectrum of $\Psi$ in presence of $\Delta$. The mass ordering among charge and neutral states is subject to alert depending on the parameters $v_\Delta$ and $y_\psi$. (b) The mass splitting forbids the diagonal neutral current. }
    \label{fig:dmy}
\end{figure}
\noindent 
Here $\delta m$ is the mass splitting between $\psi^\pm$ and $\psi^0$ and is generated due to the quantum correction as previously stated in eq.\eqref{eq:qc}.
The mass splitting between the two pseudo-Dirac states is defined by ${\delta m}_{12}=M_{\psi_2}-M_{\psi_1}=2 y_\psi v_\Delta$.
A typical mass spectrum of the dark fermion doublet is displayed in Fig-\ref{c} for the case $y_\psi v_\Delta < \delta m$ where $M_{\psi^\pm} > M_{\psi_2} >M_{\psi_1}$. 
The mass spectrum can also be  $M_{\psi_2} > M_{\psi_\pm} > M_{\psi_1}$ when $y_\psi v_\Delta > \delta m$. 
In both cases,
the lightest neutral pseudo-Dirac state  $\psi_1$ acts as a stable DM candidate. 
However, for our analysis, we adopt the former hierarchy.  
With the help of pseudo-Dirac splitting, the $Z$ mediated neutral current interaction turns into 
\bea
\overline{\psi^0} \gamma^\mu Z_\mu \psi^0 \rightarrow \psi_1 \gamma^\mu Z_\mu \psi_2 , 
\label{eq:offdd}
\eea
and prohibits to write $\psi_1 \gamma^\mu Z_\mu \psi_1$ interaction (see in eq.\eqref{eq:zint}) as shown in Fig-\ref{d}.
The advantage of this setup is that it forbids $Z$ mediated DM-nucleon elastic scattering of $\psi_1$  and helps to evade the direct detection constraint.  
We will discuss  in detail the direct detection constraints for pseudo-Dirac DM $\psi_1$ later.

\begin{figure}[!tbh]
    \centering
    \subfigure[\label{inb}]{
     \includegraphics[scale=0.34]{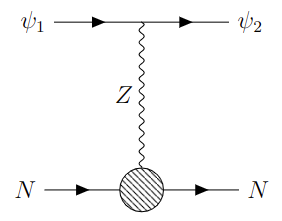}}
     \subfigure[\label{ina}]{
    \includegraphics[scale=0.3]{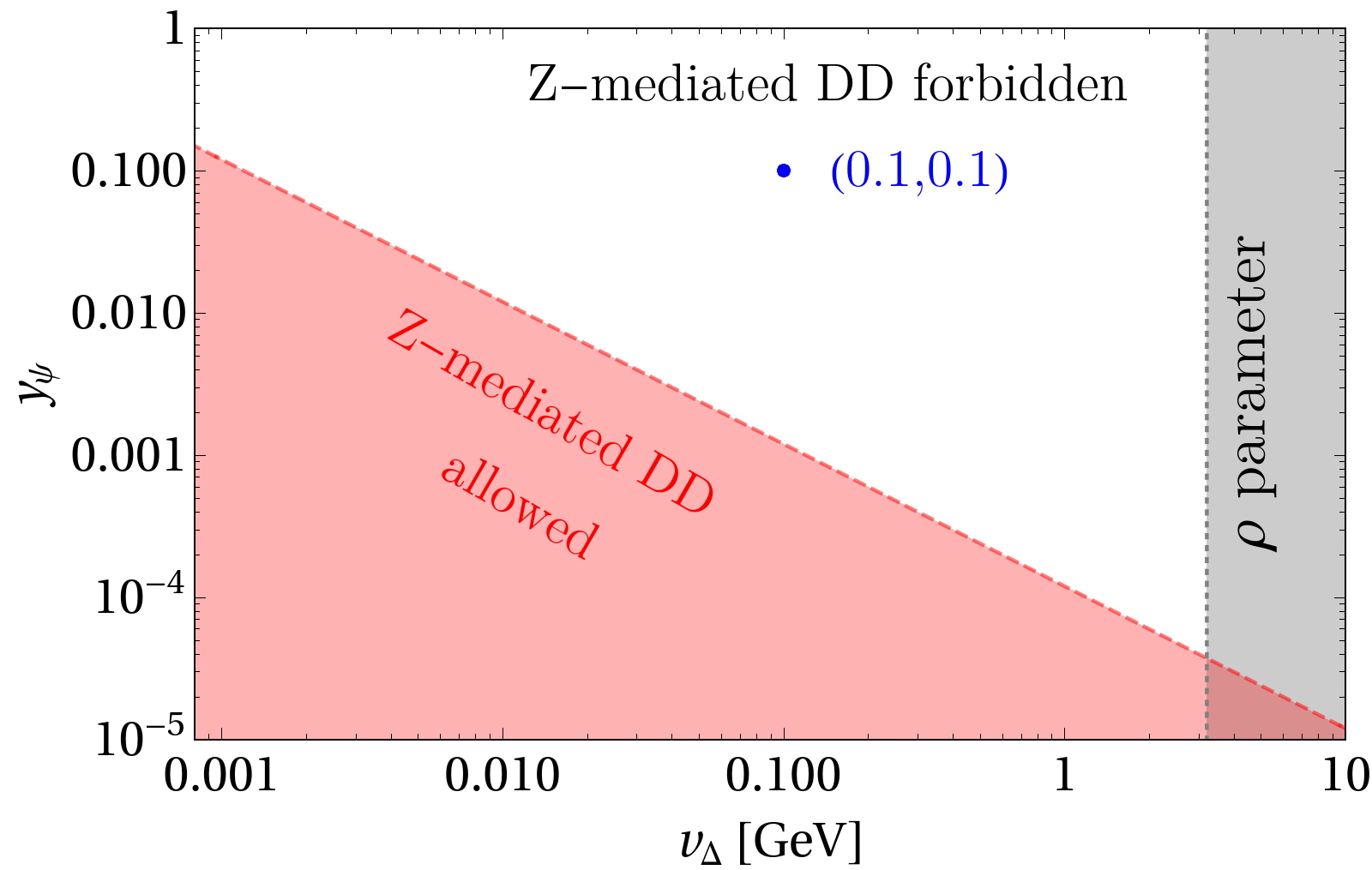}}
    \caption{\it (a)The feynman diagram for $Z$ mediated in-elastic scattering between $\psi_1$ and nucleon. (b)Allowed parameter space in the $v_{\Delta}$ vs. $y_{\psi}$ plane. The red shaded region where $Z$-mediated direct detection is allowed. The gray-shaded vertical region is excluded from $\rho$ parameter. The white region shows the choices of $y_\psi$ and $v_\Delta$ where $Z$-mediated direct detection limits on lepton doublet DM can be evaded. For our discussion, we chose a benchmark point (BP) which is shown by a blue dot with $y_\psi=0.1$ and $v_\Delta=0.1$ GeV. }
    \label{fig:indd}
\end{figure}
%===================

\noindent However, the off-diagonal $Z$ interaction with pseudo-Dirac states,$\psi_{1,2}$ in eq.\eqref{eq:offdd} allows for the inelastic DM-nucleon scattering, as illustrated in Fig-\ref{inb}.
In principle, such type of in-elastic scattering kinematically may enable direct search for extremely low mass splitting between two pseudo-Dirac states($\psi_{1,2}$), and the condition for such scenario is given as\cite{Tucker-Smith:2001myb},
\bea
({\delta m}_{12})^{\rm max.} < \frac{\beta^2}{2}\frac{M_{\psi_1} M_N}{M_{\psi_1}+M_N}
\label{eq:indd}
\eea
where $\beta c =v_{\rm DM} \simeq 650 {\rm km/s}$ (escape velocity) and $M_N$ is the nucleus mass. For the XENON 1T experiment, $Xe$ is the target nucleus with $M_N=130$ amu. Following eq.\eqref{eq:indd}, we can derive an upper limit on $({\delta m}_{12})^{\rm max.}$ as a function of DM mass $M_{\psi_1}$. This upper limit sets the threshold  below which the $Z$ mediated in-elastic scattering for direct detection is allowed. For DM mass $\sim \mathcal{O}(1  {\rm TeV})$, $({\delta m}_{12})^{\rm max.}$ turns out to be $\sim 250$ keV followed by the eq.\eqref{eq:indd}.

\noindent Therefore the Z-mediated in-elastic scattering of direct search is forbidden with ${\delta m}_{12} \gtrsim 250$ keV for $M_{\psi_1}\sim 1$ TeV. A detailed discussion on it can be found in the references\cite{Tucker-Smith:2001myb,Barman:2019tuo}. From the definition of ${\delta m}_{12}$, we can have a relation between $y_\psi$ and $v_\Delta$ which is depicted by the red dotted line in Fig.\ref{ina}. The red shaded region($2y_\psi v_\Delta < ({\delta m}_{12})^{\rm max.}$) below the red dotted line is where inelastic scattering gets allowed. The region above the red dotted line where the Z-mediated inelastic scattering is disallowed. 
Again the $\rho$ parameter puts an upper bound on $v_\Delta$ and excluded the region with $v_\Delta > 2.6$ GeV, shown by the Gray vertical region. 
Therefore our point of interest lies within the white region where $Z$ mediated in-elastic scattering is prohibited.

We shall now move to the DM phenomenology of the pseudo-Dirac DM,$\psi_1$ in the modified scenario with an added scalar triplet $\Delta$. 
A brief discussion about the interaction Lagrangian of the pseudo-Dirac DM is shown in appendix-\ref{sec:pseud}. 
The thermal abundance of DM $\psi_1$ is determined by its scalar and gauged mediated annihilation to SM ($\psi_1~\psi_1\to{\rm SM~SM}$) and co-annihilation to SM ($\psi_i~\psi_j ; \psi_i~\psi^\pm;~\psi^+~\psi^- ~\to {\rm SM~SM}$ with $i,j=1,2$). The triplet mediated diagrams depend on the Yukawa coupling $y_\psi$, the masses of the triplet scalars $M_X$, vev of the triplet $v_\Delta$ and the mixing angle between the CP even scalars $\sin\alpha$.
%%*****************************************
%*********************************************
The additional interactions of $\Psi$ with the scalar triplet can impact the relic abundance. When the masses of the triplet scalars are smaller than $M_{\psi_1}$, the new (co-)annihilation channels, $\psi_i~\psi_j(\psi^\pm); \psi^+ \psi^- \to X X~(X\equiv H, A^0, H^\pm,H^{++})$ open up and relic density of $\psi_1$ drops \cite{Bhattacharya:2018fus}. 
However in our discussion we assume $M_{\psi_1} < M_X/2$ and small $\sin\alpha$ which suppresses the additional interactions due the scalar triplet. 
Therefore in our analysis we choose a benchmark point(BP) $\sin\alpha=10^{-4},~M_\Delta=3$ TeV and $v_\Delta=0.1$ GeV and the masses of the heavy physical states ($H^0,A^0, H^\pm,H^{++}$) turns out of the order of $M_\Delta$(see eq.\eqref{eq:phystate}). 
Hence the additional scalar $\Delta$ has no significant role in deciding the relic abundance of DM.

%****************************************************************
%****************************************************
To evaluate the freeze-out abundances of DM ($\psi_1$) with $M_{\Psi} < M_{S} < M_{\Delta}/2$ for this setup, we need to solve the coupled Boltzmann equations provided in eq.\eqref{eq:cbeq}. 
And the effective thermal average cross-section, $\left< \sigma v\right>_\Psi^{\rm{eff}}$ in eq.\eqref{eq:cbeq} modified as\cite{Griest:1990kh,Edsjo:1997bg}:
\begin{widetext}
\bea
\left< \sigma v\right>_\Psi^{\rm{eff}}&=& \frac{g_{\psi_1}^2}{g_{\rm eff}^2} \left< \sigma v \right>_{\psi_1\psi_1} + \frac{2 g_{\psi_1} g_{\psi_2}}{g_{\rm eff}^2} \left< \sigma v \right>_{\psi_1\psi_2}(1+\delta_{\psi_2})^{\frac{3}{2}}~ e^{-\xi \delta_{\psi_2}} +  \frac{2 g_{\psi_1} g_{\psi^\pm}}{g_{\rm eff}^2} \left< \sigma v \right>_{\psi_1\psi^\pm}(1+\delta_{\psi^\pm})^{\frac{3}{2}}~ e^{-\xi \delta_{\psi^\pm}}\nonumber \\
&& +  \frac{2 g_{\psi_2} g_{\psi^\pm}}{g_{\rm eff}^2} \left< \sigma v \right>_{\psi_2\psi^\pm}(1+\delta_{\psi_2})^{\frac{3}{2}}~(1+\delta_{\psi^\pm})^{\frac{3}{2}}~ e^{-\xi \left( \delta_{\psi_2}+\delta_{\psi^\pm} \right)}\nonumber \\
&& +  \frac{g_{\psi_2}^2}{g_{\rm eff}^2} \left< \sigma v \right>_{\psi_2\psi_2}(1+\delta_{\psi_2})^3~ e^{-2 \xi \delta_{\psi_2}}  +  \frac{g_{\psi^\pm}^2}{g_{\rm eff}^2} \left< \sigma v \right>_{\psi^+ \psi^-}(1+\delta_{\psi^\pm})^3~ e^{-2\xi \delta_{\psi^\pm}}~, \nonumber \\
 {\rm with}~~g_{\rm eff} &=& g_{\psi_1}+g_{\psi_2}(1+\delta_{\psi_2})~ e^{-\xi \delta_{\psi_2}}+g_{\psi^\pm}(1+\delta_{\psi^\pm})~ e^{-\xi \delta_{\psi^\pm}}~ {\rm and}~ \xi=\frac{M_{\psi_1}}{T}.
\eea
\end{widetext}
Here $\delta_{\psi_2}=\frac{M_{\psi_2}-M_{\psi_1}}{M_{\psi_1}}$ and $\delta_{\psi^\pm}=\frac{M_{\psi^\pm}-M_{\psi_1}}{M_{\psi_1}}$. The internal degrees of freedom  $g_{\psi_{1,2}}$ and $g_{\psi^\pm}$ are associated with the  dark fermion states, $\psi_{1,2}$ and $\psi^\pm$ respectively.
Due to the Yukawa coupling of $S$ with $\psi_1,\psi_2$ and $\psi^\pm$; the late decay of $S$ will give rise to non-zero abundances of all dark particles $\psi_1,\psi_2$ and $\psi^\pm$. That $ \psi^\pm$ will eventually decay to the lightest $\psi_1$ as mentioned in the context of Dirac DM. On the other hand $\psi_2$ will also decay to $\psi_1$ promptly due to the strong off diagonal neutral current ($\psi_2 \to \psi_1 Z^*\to \psi_1 \nu \Bar{\nu} $). So at late time $Y_\Psi^{x\to \infty}\simeq Y_{\psi_1}^{x\to \infty}\simeq Y_{\psi_1}(x^{\rm FO}_{\psi_1})+Y_{S}(x^{\rm FO}_{S})$ .

\begin{figure}[tbh]
    \centering
    \includegraphics[scale=0.4]{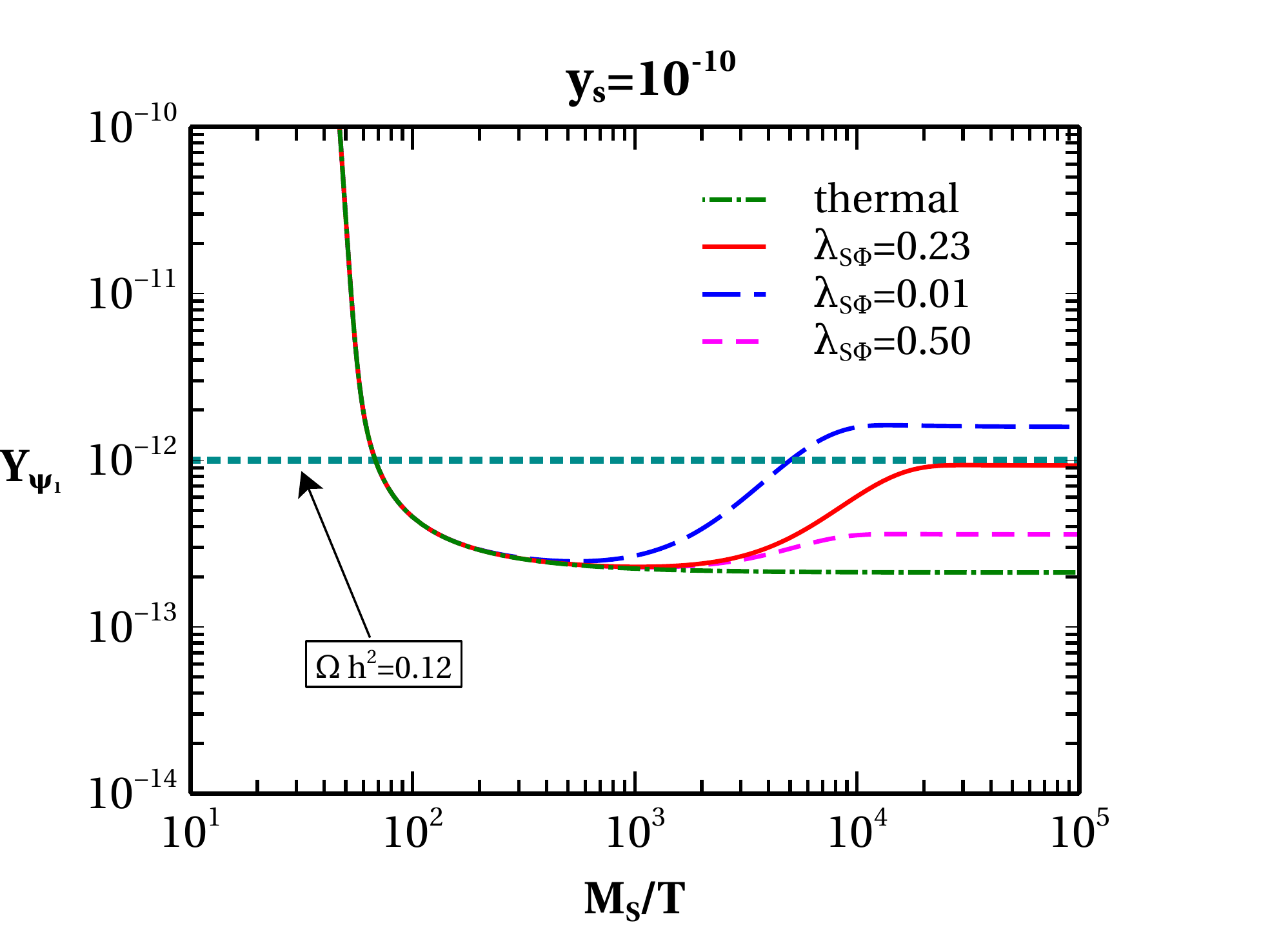}
    \caption{\it Comoving abundances of $\psi_1$ for $M_{S}=1000$ GeV, $M_{\Psi}=450$ GeV with varying $\lambda_{S \Phi}\in(0.01,0.23,0.5)$ shown by the blue dashed, red solid and magenta dashed lines respectively.
     The green dashed dot line correspond to $Y_{\psi_1}$ in absence of the dark Yukawa coupling i.e. $y_s=0$. The thick cyan dotted line corresponds to the comoving abundance leading to observed relic density for $M_{\Psi}=450$ GeV. }
    \label{fig:pseud}
\end{figure}

In Fig.-\ref{fig:pseud} we plot the comoving abundance of DM, $ Y_{\psi_1}$ with $M_S/T$ for the modified scenario. For the plot, we kept fixed $M_S=1000$ GeV, $M_{\Psi}=450$ GeV, $M_{\Delta}=3$ TeV and $y_{S}=10^{-10}$. We also kept fixed $y_\psi=0.1$ and $v_
\Delta=0.1$ GeV, resulting in pseudo-Dirac mass splitting ${\delta m}_{12}=20$ MeV,  consistent with the direct detection requirement. 
The green dot-dashed line denotes the abundance of DM in the absence of $S$ i.e. no late-time DM production.
The cyan dashed line corresponds to the comoving density leading to the correct relic density measured by PLANCK for $M_{\Psi}=450$ GeV .
The blue dashed, red solid and magenta dashed lines correspond to the three different values of $\lambda_{S \Phi}=0.01,0.23$ and $0.5$ respectively.
From the figure we notice that as the $\lambda_{S\Phi}$ increases, $Y_\psi(\simeq Y_{\psi_1})$ decreases. 
With increase in $\lambda_{S\Phi}$, the annihilation cross-section of $S$ to SM increases leading to lower freeze out abundance of $S$, $Y_S^{x_F}$. The freeze-out density of $S$ completely dilutes into the $Y_{\psi_1}$ through the late-time decay of $S$ to $\psi_1$, and the abundance of DM eventually added up as: $ Y^{x\to \infty}_{\psi_1} \equiv  Y_{\psi_1}(x^{\rm FO}_{\psi_1})+Y_{S}(x^{\rm FO}_{S})$. 
 Therefore lower $Y_S({x_S^{\rm FO}})$ causes lower $Y^{x\to \infty}_{\psi_1}$. This property, like the Dirac DM, is easily portrayed in the above-mentioned figure.
 It is important to note that in the modified scenario, the small pseudo-Dirac mass splitting ${\delta m}_{12} \ll M_{\psi_1}$  has no effect on the DM abundance and that DM abundance is almost the same as the Dirac like scenario.  As a result, the study of DM relic density under the circumstances of ${\delta m}_{12} \ll M_{\psi_1}$ and $M_{\Psi,S}< M_{\Delta}/2$  remains altered for both the scenarios. 
  However, direct detection distinguishes both the Dirac and pseudo-Dirac DM cases as discussed earlier. We now discuss the experimental constraints for the pseudo-Dirac doublet DM scenario.
%===========================================
\subsection{Direct Detection}
\begin{figure}[!tbh]
    \centering
    \subfigure[\label{dd_a}]{
    \includegraphics[scale=0.37]{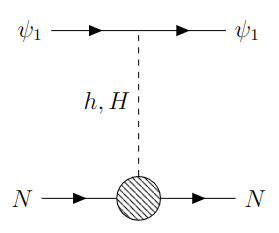}}
    \subfigure[\label{dd_b}]{
    \includegraphics[scale=0.37]{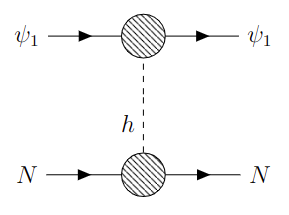}}
    \caption{\it Elastic scattering of DM $\psi_1$ with nucleon via (a) tree level and (b) loop level process.}
    \label{fig:p_DD}
\end{figure}

%===========================================
We discussed in the context of Dirac DM that current bounds from the direct detection experiment rule out Dirac doublet DM for mass even up to 2 TeV. The Majorana nature of the pseudo-Dirac DM  easily evades the direct detection constraint. In the modified scenario, the presence of the triplet scalar, which is responsible for pseudo-Dirac splitting,  determines the fate of this model in the direct search experiment. Therefore the DM particles can recoil against the nucleus,  producing the direct search signature (spin independent) via both tree and loop-level DM-nucleon 
scattering processes as displayed  in Fig.\ref{fig:p_DD}.
%====================
%=================
In Fig.\ref{fig:p_DD} we show the Feynman diagrams for tree-level processes via scalar mixing and loop-level processes via gauge bosons which impact significantly in direct detection cross-section.
The loop contributions occur via the gauge mediated diagrams, shown in the appendix 
\ref{sec:loop}, which can have a significant contribution to the elastic scattering between DM and detector nucleon.
The amplitude for direct detection will be $\mathcal{M}^{\rm SI-DD}=\mathcal{M}_{\rm tree}+\mathcal{M}_{\rm loop}$.
However, for our choice of parameters $M_\Delta=3$ TeV and $\sin\alpha \sim \mathcal{O}(10^{-4})$, the tree level amplitude is mixing and propagator suppressed, while the gauge mediated loop induced amplitude is dominating.
 A brief discussion on spin-independent direct detection cross-section of the DM $\psi_1$ followed from \cite{Hisano:2011cs}, can be found in appendix \ref{sec:loop}. 
We found that the loop level contribution($\sigma^{\rm SI} \sim 10^{-47}$ cm$^2$) is dominating over the tree level contribution($\sigma^{\rm SI} \sim 10^{-54}$ cm$^2$) for $M_{\psi_1}\sim \{100-1000\}$ GeV with the aforementioned benchmark parameters. Therefore the total spin-independent direct detection cross section is well below the existing bounds obtained by the different direct search experiments like XENON-1T\cite{XENON:2017vdw}, Panda-4T\cite{PandaX-II:2016vec} and LZ\cite{LZ:2022ufs}. 
%===========================================
\subsection{ Indirect Detection}
%===========================================
\begin{figure}[tbh]
    \centering
    \includegraphics[scale=0.38]{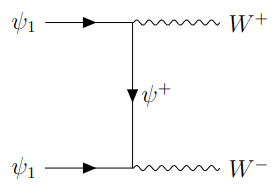}
    \caption{\it Feynman diagram corresponds to the indirect search mode:  $\psi_1\psi_1 \to W^+W^-$.  }
    \label{fig:ww1}
\end{figure}
Similar to direct detection, the relevant constraint may arise from indirect search experiments like Fermi-LAT\cite{Fermi-LAT:2016uux} and MAGIC\cite{MAGIC:2016xys} by analyzing excess gamma-ray flux. The  excess gamma-ray flux can be produced via the production of
the SM particles either through DM annihilation or via decay in the local Universe. In this scenario, DM annihilation processes  $\psi_1~\psi_1 \to X \overline{X}$ where $X=\{W^-,~b, ~\mu^-, ~\tau^-\}$; and the subsequent decay $X$ to photons resulted in the production of the gamma ray. Non-observation of DM at the indirect search experiments like Fermi-LAT and MAGIC put an upper bound on the individual thermal averaged annihilation cross-section. For the DM mass $M_{\psi_1}>M_W$, the most stringent constraint comes from the annihilation process $\psi_1\psi_1 \to W^-W^+$. The corresponding Feynman diagram is shown in Fig.\ref{fig:ww1} which is mediated by $\psi^\pm$.    

\begin{figure}[!tbh]
    \centering
    \includegraphics[scale=0.4]{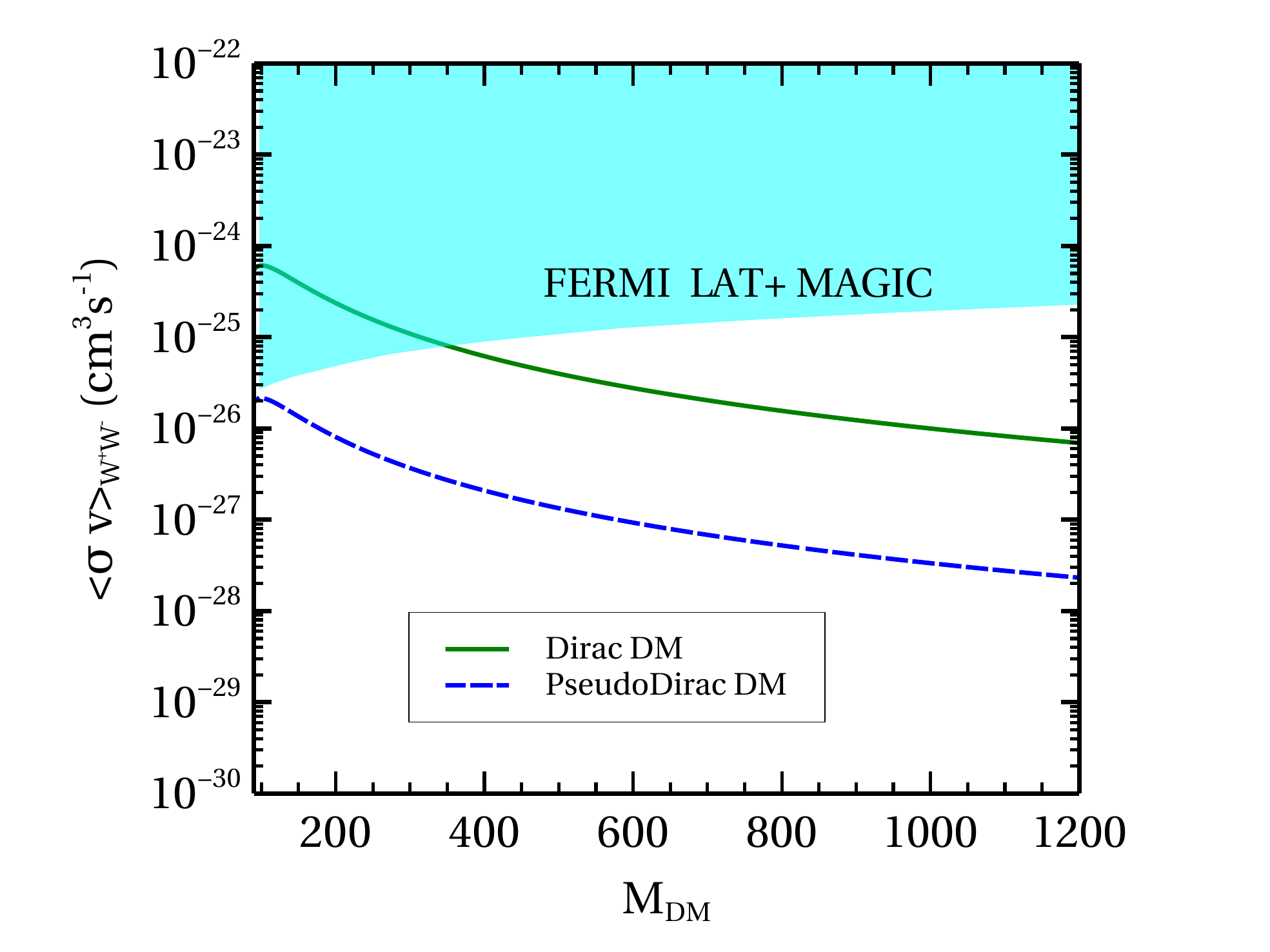}
    \caption{\it Indirect detection cross-section for  DM annihilation to $W^+W^-$ as a function of DM mass. The blue dashed line corresponds to the pseudo-Dirac DM scenario where DM is ${\psi_1}$ and the green solid line represents the Dirac DM where DM is ${\psi^0}$. The combined exclusion bound from Fermi LAT and MAGIC for the DM annihilation channel is shown by cyan region in the same plane.}
    \label{fig:ww}
\end{figure}

Note the $\left<\sigma v\right>_{_{W^+W^-}}$
for pseudo-Dirac DM $\psi_1$ 
is smaller than for Dirac DM $\psi^0$.
The reason for this is the absence of $Z$ mediated $s$-channel diagram for the process $\psi_1\psi_1 \to W^+W^-$ in the case of pseudo-Dirac DM.
In Fig.\ref{fig:ww}, we plot $\left<\sigma v\right>_{_{W^+W^-}}$ as a function of DM mass along with the combined Fermi-LAT and MAGIC exclusion bound. It turns out that 
the $ \left<\sigma v\right>_{_{W^+W^-}}$ lies below the bound from indirect detection. Note for the Dirac doublet DM set up the same annihilation channel excluded the DM mass below $\sim 350$ GeV.
%===========================================
\subsection{Collider constraint}
%===========================================
As stated earlier, due to quantum correction, there exists tiny mass splitting ($\delta m \sim m_\pi$) between the charged and neutral component of the doublet. For such tiny mass splitting, the dominate decay mode of the charged fermion is $\psi^\pm \to \psi_1 \pi^\pm$. The corresponding decay width is given by,
%\bew
\bea
\Gamma (\psi^\pm \to \pi^\pm \psi_1) 
&\approx& \frac{G_F^2 (f_\pi \cos\theta_c)^2 }{2\pi} ~ \left( \Delta m \right)^3 \nonumber\\
&& \times \Bigg(1-\frac{m^2_\pi}{\left( \Delta m\right)^2} \Bigg)^{1/2}  
\propto \left( \Delta m\right)^3 ,\nonumber
\label{eq:decypsip}
\eea
%\eew
where $\Delta m$ is the mass splitting between $\psi^\pm$ and $\psi_1$, defined as $\Delta m=\delta m+ {\delta m}_{12}$.
Here $G_F=1.16638\times10^{-5}$, $\sin\theta_c=0.22$, $f_\pi=130$ MeV and $m_\pi=139.57$ MeV.   
If the final decay product is low momentum $\pi^{\pm}$ and stable $\psi_1$,
it leaves the detector without interaction.
Such signature is called the displaced vertex (DV) signatures and LHC has already been constrained such scenario in the context of Higgsino \cite{Calibbi:2018fqf,Belyaev:2020wok}.
As Higgsino has a similar setup to our doublet fermion DM, we adopt their bound and it turns out for $M_{\psi_1} \lesssim 450$ GeV is  excluded from the displaced searches at 8 TeV LHC\cite{Calibbi:2018fqf}.
%===========================================
\subsection{Results}
\begin{figure}[!tbh]
    \centering
    \includegraphics[scale=0.4]{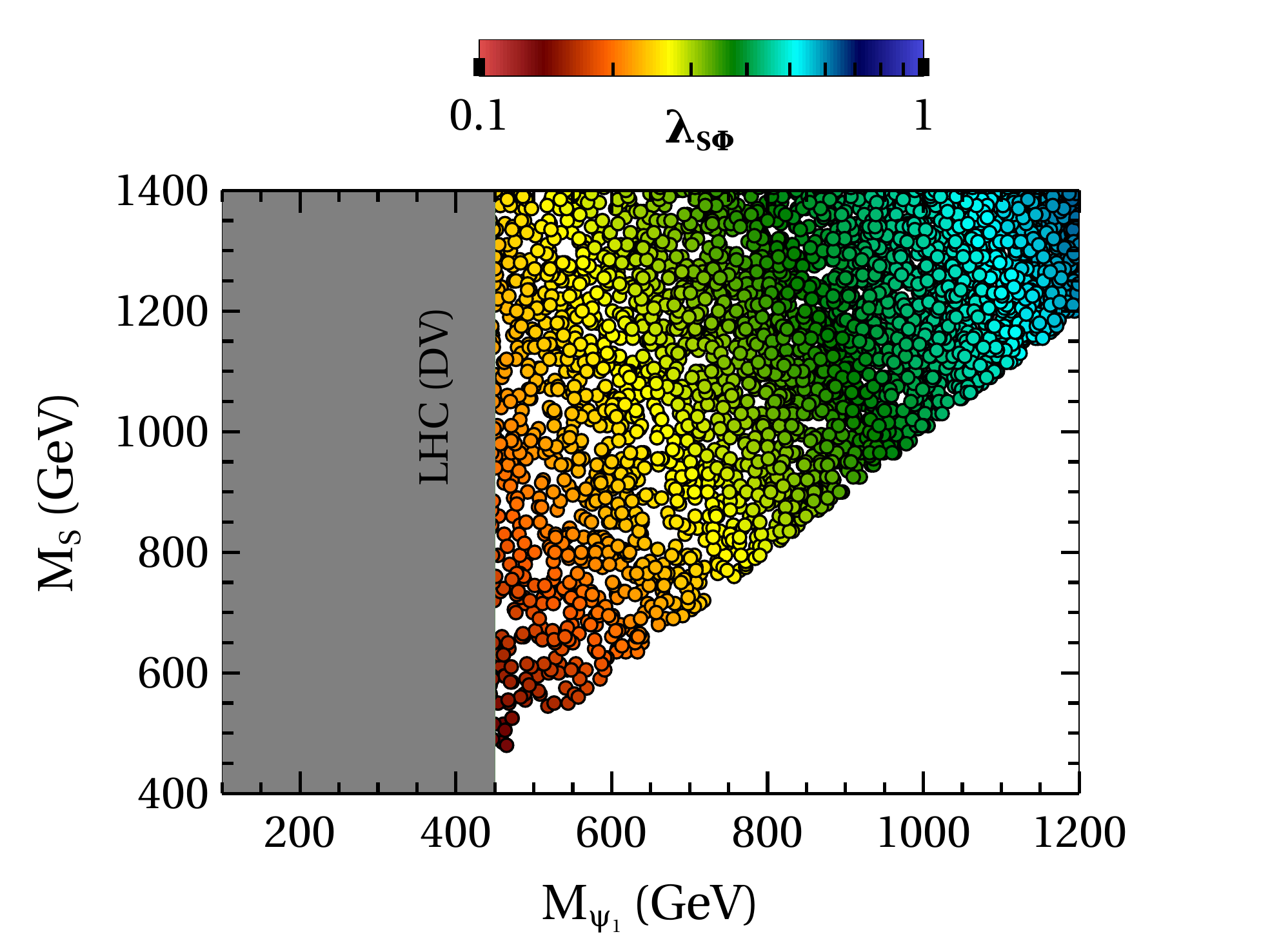}
    \caption{\it Points satisfying observed relic density in $M_{\psi_1}$ vs. $M_S$ plane. The color bar resembles the variation in $\lambda_{S \Phi}\in \{0.1- 1.0\}$.  The grey region is excluded from LHC displaced vertex search. The white region represents the parameter space where the lepton doublet can not be stable.}
    \label{fig:scan}
\end{figure}

%===========================================
Finally, we show all the parameter space satisfying the observed relic density via our proposed hybrid setup.
In Fig.-\ref{fig:scan} we do numerical scan with varying $M_{\psi_1},~M_S,~\lambda_{S\Phi}$ for a fixed Yukawa coupling $y_s=10^{-10}$. 
We took a BP with $M_{\psi_2}-M_{\psi_1}=20$ MeV, $\sin{\alpha}=10^{-4},~v_{\Delta}=0.1$ GeV and the masses of the heavy scalars ($m_{H_2,A^0,H^{+},H^{++}}\sim 3$ TeV) such that they have no effect on DM abundance. 
With such parameter choices, the scalar triplet sector has no such role in deciding $Y_{\psi_1}({x_{\psi_1}^{\rm FO}})$ apart from evading the stringent direct detection bound.
Similar to Dirac DM scenario for $M_{\psi_1}<1200$ GeV, $Y_{\psi_1}^{x \to \infty}$ is decided by the gauge interactions and the late time decay of $S$ repopulate $Y_{\psi_1}$ leading to observed relic density.
In the above mentioned plot we varied $M_{\psi_1}$ upto $1200$ GeV and $M_{S}$ upto $1400$ GeV maintaining  $M_S>M_{\psi_1}$
GeV so that $S$ can decay on-shell.
The color variation in the plot depicts the variation in $\lambda_{S\Phi}\in \{0.1-1\}$.
From the figure, we notice that for fixed $M_S$, increase in $M_{\psi_1}$ calls for an increase in $\lambda_{S\Phi}$.
The reason for such a feature is the fact that  a higher value of $M_{\psi_1}$ leads to a higher $Y_{\psi_1}^{x_F}$ and to meet the observed relic from late-time decay of $S$, lower value of $Y_{S}^{x_F}$ is required. 
As $Y_{S}^{x_F}$ is inversely proportional to $\lambda_{S\Phi}$ for higher $M_{\psi_1}$, the higher value of $\lambda_{S\Phi}$ is needed. This feature is elaborated in the context of Fig.-\ref{fig:pseud}.
The grey region is excluded from the displaced vertex (DV) signatures\cite{Belyaev:2020wok} which gives the most stringent bound among the experimental constraints. 
%==============================%=============
\section{Conclusion}
\label{sec:conc}
%============================================
In this work, we study a minimal setup for lepton doublet dark matter satisfying the observed relic density and direct search constraints in the sub-TeV mass range. The dark matter relic is produced in two different epochs.
At first DM abundance is generated thermally from bath particles. At the later epoch, an additional dark sector particle contributes non-thermally to accommodate the total observed DM abundance. The additional dark sector particle was initially in the thermal bath, but it decoupled at the time of non-thermal DM production. This hybrid setup differs from pure WIMP and pure FIMP instances. To realise the scenario, we extend the SM particle content by a $SU(2)_L$ lepton doublet $\Psi$ and a SM gauge singlet scalar $S$.
An additional $\mathcal{Z}_2$ symmetry has been imposed under which both $\Psi$ and $S$ are odd, while all the SM particles are even. In the mass hierarchy $M_\Psi < M_S$, the neutral component of the lepton doublet, $\psi^0$ serves as a stable DM candidate. 
 The additional DM production from the late-time non-thermal production can help to compensate for the under-abundance problem in the sub-TeV mass region of $\Psi$. However $\psi^0$ still faces challenges from direct detection constraints in the sub-TeV mass ranges due to its strong gauge-mediated interactions. 
 To overcome the severe constraint of direct detection (DD), we add a $\mathcal{Z}_2$ even scalar triplet $\Delta$ ($Y_\Delta =2$) which helps to generate pseudo-Dirac nature of DM($\psi_1$), leading to non-diagonal neutral current gauge interaction. Thus using this setup the DM $\psi_1$ can evade the direct detection constraint.  
Apart from satisfying the observed DM abundance within the mass range, $M_{\psi_1} \lesssim 1200$ GeV, the model
has promising detection prospects at the direct, indirect, and collider search experiments. 
The most stringent limit comes from displaced vertex (DV) signatures\cite{Belyaev:2020wok}, which
 excludes DM masses less than $450$ GeV. As a result, the DM mass is consistent with all the constraints within the mass range $M_{\psi_1} \sim 450-1200$ GeV. The DM mass $M_{\psi_1}\gtrsim 1200$
GeV is ruled out from relic density constraint where DM becomes overabundant. Apart from explaining sub-TeV lepton doublet DM, this minimal framework can also address the non-zero neutrino masses and mixing in presence of the scalar triplet. 
 Thus we present an alternate mechanism where such kind of hybrid set up can provide additional DM abundance. This can help to revive a substantial part of parameter space where the minimal DM models fail to fulfill the observed relic density like in electroweak DM multiplet models.
%=========================================
\section*{Acknowledgement}
The authors thank Dilip Kumar Ghosh for the insightful discussion and helpful suggestions.
SJ is supported by CSIR, Government of India, under the NET JRF fellowship
scheme with Award file No. 09/080(1172)/2020-EMR-I.

%=================================
%\clearpage
\appendix
%============================================
\section{Interactions of Dirac doublet}
\label{sec:apxdirac}
The relevant Lagrangian($\mathcal{L}_{\Psi}$) containing the interactions of Dirac doublet is given by
\bea
\mathcal{L}_\Psi^{\rm int}&=&\frac{e}{\sin 2\theta_w } \overline{\psi^0}
\gamma^{\mu} Z_{\mu} \psi^0 - e \cot 2\theta_w {\psi^+}
\gamma^{\mu} Z_{\mu} \psi^- \nonumber 
\\&& 
+\frac{e}{\sqrt{2} \sin \theta_w} \left(\overline{\psi^0} \gamma^{\mu} W_{\mu}^+ \psi^- +h.c \right) \nonumber \\
&& -e~ \psi^+ \gamma^{\mu}A_{\mu} \psi^-, 
\eea
where $e$ is the electromagnetic coupling
constant and $\theta_w$ is the Weinberg angle.
%=========================================
%%%%%%%%%%%%%%%%%%%%%%%%%
\section{A brief description of the scalar sector }
\label{sec:apxDelta}
%================
 The Lagrangian for the scalar sector containing a SM like Higgs doublet $\Phi$ and a scalar triplet $\Delta$ with hyper-charge $Y_\Delta = 2$ reads as\cite{Ghosh:2022fzp}
\bea
\mathcal{L}_{\Delta + \rm SM}&=&\left(D^{\mu}\Phi\right)^{\dagger}\left(D_{\mu}\Phi\right) + {\rm Tr} \left[\left(D^{\mu}\Delta\right)^{\dagger}\left(D_{\mu}\Delta\right)\right] \nonumber \\
&&
-V(\Delta,\Phi)-y_{L} \bar{L^c}i \sigma^2\Delta L+h.c. ~,
\label{eq:typeII}
\eea
where the definition of the co-variant derivative of both the doublet and triplet scalar is defined as,
\bea
D_{\mu}\Phi &=& \Big(\partial_\mu - i g_2 \frac{\sigma^a}{2} W_{\mu}^a-i g_1 \frac{Y_{\Phi}}{2} B_{\mu}\Big)\Phi , \nonumber \\
D_{\mu}\Delta &=& \partial_{\mu}\Delta - i g_2 \left[\frac{\sigma^{a}}{2} W_{\mu}^a ,\Delta\right] - {i g_1} \frac{Y_\Delta}{2} B_{\mu}\Delta~.
\eea
The most general scalar potential including $\Delta$ and $H$ can be written as: 
\bea
V(\Delta,\Phi) &=&  -\mu_{\Phi}^2 (\Phi^\dagger \Phi)+ \lambda_{\Phi} (\Phi^\dagger \Phi)^2 ~ \nonumber \\
&& + \mu_{\Delta}^2 {\rm Tr}\left[\Delta^{\dagger}\Delta\right]  + \lambda_{1} \left(\Phi^{\dagger}\Phi\right){\rm Tr}\left[\Delta^{\dagger}\Delta\right] 
\nonumber \\
&& 
+ \lambda_2 \left({\rm Tr}[\Delta^{\dagger}\Delta]\right)^2 
+\lambda_3 ~{\rm Tr} [\left(\Delta^{\dagger}\Delta\right)^2]
\nonumber \\
&&  + \lambda_4 ~\left(\Phi^{\dagger}\Delta\Delta^{\dagger}\Phi \right)+ \left[\mu_1\left(\Phi^T i \sigma^2 \Delta^{\dagger}\Phi\right)+h.c.\right]. \nonumber \\
\eea
Note $\mu_\Delta^2 >0$ and the $\Delta$ does not acquire any vev. However the cubic term in the scalar potential $\Phi^T i \sigma^2 \Delta^{\dagger}\Phi$ leads to the generation of an induce non-vanishing small vev $v_\Delta$ for the $\Delta$ after the EWSB. Then the scalar fields, $\Delta$ and $\Phi$ can be represented as: 
\bea
\Delta= \left( \begin{matrix}
                   \frac{\Delta^+}{\sqrt{2}} && \Delta^{++}   \\
                    \frac{1}{\sqrt{2}} (v_{\Delta} +\delta +i \eta)&&  -\frac{\Delta^+}{\sqrt{2}}  
                        \end{matrix}\right) ,
\Phi= \left( \begin{matrix} 
                \phi^+ \\
                \frac{1}{\sqrt{2}} (v_d + \phi +i \chi)
            \end{matrix} \right) \nonumber
\eea
In the alignment limit $v_\Delta \ll v_d$ and $v=\sqrt{v_d^2+v_\Delta^2}=246$ GeV. Minimizing the
scalar potential at the vacuums ($v_d$ and $v_\Delta$) leads to the following conditions
\bea
\mu_H^2 &=& -\frac{2 M_\Delta^2 v_{\Delta}^2}{v_d^2}+\frac{1}{2} v_{\Delta}^2 (\lambda _1+\lambda _4)+\lambda_H v_d^2  \nonumber \\
\mu_\Delta^2 &=& M_\Delta^2-v_{\Delta}^2 (\lambda_2+\lambda_3)-\frac{1}{2} v_d^2 (\lambda_1+\lambda_4)
\eea
with $M_{\Delta}^2=\frac{\mu_1  v_d^2}{\sqrt{2} v_{\Delta}}.$

\noindent
After the EWSB, the two CP even states $\phi$ and $\delta$ mixed up. The mass matrix can be diagonalized using the orthogonal rotation followed by 
\bea
\left( \begin{matrix}
                \phi \\
                \delta
          \end{matrix} \right) =
 \left( \begin{matrix} 
                   \cos{\alpha} && \sin{\alpha}   \\
                    -\sin{\alpha} && \cos{\alpha} 
                        \end{matrix} \right)   
\left( \begin{matrix}
                h \\
                H
          \end{matrix} \right) .                       
\eea
with the mixing angle
\bea
\tan{2\alpha}= \frac{v_{\Delta}}{v_d} \frac{2(\lambda_3 +\lambda_4)v_d^2 -4 M_{\Delta}^2}{(-2 \lambda_H v_d^2 +4(\lambda_1 +\lambda_2)v_{\Delta}^2 +M_{\Delta}^2)}
\eea
The above orthogonal transformation gives rise to two physical states $h$ and $H$ with the physical masses $m_{h} (\simeq 125.09$ GeV) and $m_H$ respectively.
Similarly, the CP-odd states are mixed up and lead to one massless Goldstone state eaten by a massive SM $Z$ boson and a massive CP odd eigen state $A^0$ with mass $m_{A^0}$. The orthogonal rotation of the singly charge scalars $\phi^+$ and $\Delta^+$ yields one massless Goldstone mode, absorbed by SM $W$ boson and one massive charged eigen state $H^\pm$ with mass $m_{H^\pm}$. The scalar sector also has one massive doubly charged eigen state $H^\pm (\equiv \Delta^\pm)$ with mass $m_{H^\pm}$.     

\noindent
The masses of the physical scalars are defined in terms of the couplings, vevs, mixing angle $\alpha$, and the new mass parameter $M_\Delta$ as:  
%=========
\bea
m_{h}^2&=& \left(M_\Delta^2+2 v_{\Delta}^2 (\lambda_2+\lambda_3)\right) \sin ^2\alpha + 2 \lambda_H v_d^2 \cos ^2\alpha
\nonumber \\
%=====
&&-\frac{v_{\Delta} \sin2\alpha \left(2 M_\Delta^2 - v_d^2 (\lambda_1+\lambda_4)\right)}{v_d}   \nonumber \\
%===
m_{H}^2&=& \left(M_\Delta^2+2 v_{\Delta}^2 (\lambda_2+\lambda_3)\right) \cos ^2\alpha + 2 \lambda_H v_d^2 \sin^2\alpha
\nonumber \\
%=====
&& +\frac{v_{\Delta} \sin2\alpha \left(2 M_\Delta^2 - v_d^2 (\lambda_1+\lambda_4)\right)}{v_d} \nonumber\\
%===
m_{A^0}^2&=& \frac{M_\Delta^2 \left(4 v_{\Delta}^2+v_d^2\right)}{v_d^2} \nonumber\\
%===
m_{H^\pm}^2 &=& \frac{\left(2 v_{\Delta}^2+v_d^2\right) \left(4 M_\Delta^2 - \lambda_4 v_d^2\right)}{4 v_d^2} \nonumber \\
%===
m_{H^{\pm\pm}}^2 &=& M_\Delta^2-\lambda_3 v_{\Delta}^2-\frac{\lambda_4 v_d^2}{2} .
\label{eq:phystate}
\eea

The Yukawa interaction involving SM lepton and $\Delta$ in eq.\eqref{eq:typeII}, can generate light neutrino masses via Type-II seesaw mechanism \cite{Cheng:1980qt,Mohapatra:1980yp}.

%%%%%%%%%%%%%%%%%%%%%%%%%%%%%%%%%%%%%%%%%%
\section{Interactions of pseudo-Dirac doublet}
\label{sec:pseud}
The mass matrix for the neutral lepton in the basis $(\psi^0 ~{\psi^0}^c)^T$ : 
\begin{eqnarray*}
\mathcal{L}^{\rm Pseudo-Dirac}_{\rm mass}&=&  
\frac{1}{2} \overline{\left( \begin{matrix} 
                  \psi^0 && {\psi^0}^c   \end{matrix} \right)}
\left( \begin{matrix} 
                  M_\Psi  && y_{_\psi} v_{_\Delta}    \\
                     y_{_\psi} v_{_\Delta} && M_\Psi   
                        \end{matrix} \right)
\left( \begin{matrix} 
                  \psi^0 \\ {\psi^0}^c   \end{matrix} \right)\\
%%%% 
&=&\frac{1}{2} \overline{\left( \begin{matrix} 
                  \psi_1 && \psi_2   \end{matrix} \right)}
\left( \begin{matrix} 
                  M_{\psi_1} && 0   \\
                   0 && M_{\psi_2}  
                        \end{matrix} \right)
\left( \begin{matrix} 
                  \psi_1 \\ \psi_2   \end{matrix} \right),
\end{eqnarray*}
where $\psi_1$ and $\psi_2$ two pseudo-Dirac states with mass $M_{\psi_1}$ and $M_{\psi_2}$ respectively and expressed as
\bea
\psi_1&=&\frac{i}{\sqrt2}({\psi^0}^c-\psi^0)~{\rm with}~M_{\psi_1}= \left( M_\Psi -  v_{_\Delta} y_{_\psi} \right)\nonumber \\
\psi_2&=&\frac{1}{\sqrt2}({\psi^0}^c+\psi^0)~{\rm with}~M_{\psi_2}= \left( M_\Psi +  v_{_\Delta} y_{_\psi} \right) 
\label{eq:pstate}
\eea
We can translate the whole Lagrangian into the physical ($\psi_1,\psi_2$) basis. The kinetic part apart from the gauge interactions will be
\be 
\mathcal{L}_{\rm KE}= \frac{1}{2}\overline{\psi_1}\gamma^{\mu}\partial_{\mu}\psi_1+ \frac{1}{2} \overline{\psi_2}\gamma^{\mu}\partial_{\mu}\psi_2 + \overline{\psi^-}\gamma^{\mu}\partial_{\mu}\psi^- .
\ee
The interaction with only $\psi^{\pm}$ will not change but those with involving 
$\psi^0$ will change. 
The interaction of neutral current of $\psi^0$ will be 
\bea
\mathcal{L}_{\rm NC}&=& g_Z \frac{1}{\sqrt{2}}(\overline{\psi_2+i \psi_1}) \gamma^{\mu} Z_{\mu}\frac{1}{\sqrt{2}}(\psi_2+i \psi_1) \nonumber\\
&=& i~g_{_Z} \frac{1}{2} \overline{\psi_2} \gamma^{\mu} Z_{\mu} \psi_1 
-i g_{_Z} \frac{1}{2} \overline{\psi_1} \gamma^{\mu} Z_{\mu} \psi_2 \nonumber\\
&=& i~g_{_Z} \overline{\psi_2} \gamma^{\mu} Z_{\mu} \psi_1~ ,
\label{eq:zint}
\eea
where $g_{_Z}={e}/{\sin 2\theta_w }$.

\noindent
Now, we would like to discuss the most important part of the diagonal neutral current interactions, ($\overline{\psi_1} \gamma^{\mu} Z_{\mu} \psi_1$).
We note from eq.\eqref{eq:pstate} that $\psi_{1,2}^c=\psi_{1,2}$. Therefore, 
\begin{eqnarray*}
\nonumber
\overline{\psi_i} \gamma^{\mu}  \psi_k &=& \overline{\psi_i^c} C^{-1}(\gamma^{\mu})^T C \psi_k^c \\
&=&  - \overline{\psi_i} \gamma^{\mu}  \psi_k~(\text{using} C^{-1}(\gamma^{\mu})^T C= - \gamma^{\mu} ). 
\label{eq:zint1}
\end{eqnarray*}
So, for $i=k$ the above mentioned diagonal term identically becomes zero 
\cite{Akhmedov:2014kxa}.

\noindent
The interaction of the neutral component with the gauge boson $W$ will also be expressed as follows:
\bea
\mathcal{L}_{\rm CC}&=&\frac{e}{\sqrt{2}\sin \theta_w}\left(  \overline{\psi^0} \gamma^{\mu} W_{\mu}^+ \psi^- +h.c \right) \nonumber \\
&=&
\frac{e}{\sqrt{2}\sin \theta_w} \times \nonumber \\
&&\left(\frac{1}{\sqrt{2}}\overline{(\psi_1+i \psi^2)} \gamma^{\mu} W_{\mu}^+ \psi^-  +h.c \right)
\eea
%=====================================
{\bf Interactions with triplet:}\\
There will also be some new interactions with the lepton doublet $\Psi$ due to the Yukawa coupling with $\Delta$.
The relevant Lagrangian is given by,
\bea
\nonumber
 -\mathcal{L}_{\Delta-\Psi}^{\rm Yuk.}&=& \frac{y_\psi}{\sqrt{2}} \overline{\Psi^c}~i \sigma_2~\Delta \Psi  +~h.c. \nonumber \\
&=&   -\frac{y_\psi}{4} \left(\psi_1 \psi_1~-\psi_2 \psi_2\right)\left(-\sin\alpha ~h+ \cos\alpha ~H \right)\nonumber \\
&& -i\frac{y_\psi}{4}~\left(\psi_1 \psi_1~- \psi_2 \psi_2\right) A^0 \nonumber \\
&& - \frac{y_\psi}{\sqrt{2}} H^{+}\left(i\psi_1+\psi_2 \right) \psi^- \nonumber \\
&&+\frac{ y_\psi}{\sqrt{2}}  H^{++} \psi^- \psi^- +h.c. 
\label{eq:int_triplet}
\eea
%============================================
%%%%%%%%%%%%%%%%%%%%%%%
\section{Loop mediated Direct detection cross section}
\label{sec:loop}
%=============
The spin independent(SI) direct detection cross-section is calculated using the effective Lagrangian\cite{Hisano:2011cs} 
\bea
\mathcal{L}_{\rm SI-DD}^{\rm eff}&=& \sum_{q=u,d,s} \left(\lambda_q^{\rm tree}+f_q \right) m_q {\psi_1}\psi_1 \Bar{q} q  \nonumber \\
&& + f_G {\psi_1}\psi_1 G_{\mu\nu} G^{\mu\nu}.
\eea
Here $\lambda_q^{\rm tree}$ denotes the Higgs mediated tree level interaction given in Fig.\ref{dd_a} which is given by

\bea
\lambda_q^{\rm tree} =  \frac{y_\psi \sin{2\alpha}}{4 v_d} \left(  \frac{1}{m_h^2} -\frac{1}{m_H^2} \right).
\eea

 However $f_q$ is the gauge mediated($W,Z$) loop level DM-nucleon interaction and $f_G$ is the loop mediated interaction with gluons as discussed in Ref.\cite{Hisano:2011cs}. The SI DM-nucleon elastic cross-section in this case is given by 
 
 \bea
 \sigma_{\rm DM-N}&=& \frac{4}{\pi} \left( \frac{m_N M_{\psi_1}}{m_N+M_{\psi_1}}\right)^2 |f_N|^2 \\
  {\rm where} ~&& \frac{f_N}{m_N} = \sum_{q=u,d,s}  \left(\lambda_q^{\rm tree}+f_q \right) f_{Tq} -\frac{8\pi }{9\alpha_s} f_{TG} f_G . \nonumber 
 \eea
 
\begin{figure}[H]
    \centering
    \includegraphics[scale=0.37]{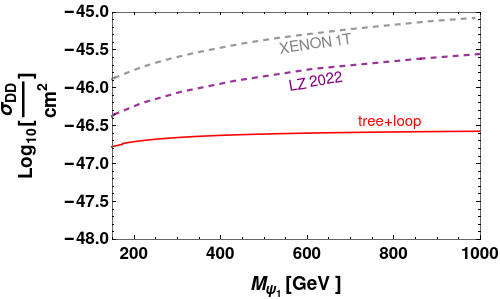}
    \caption{\it Variation of $\sigma_{\rm DD}$ with $M_{\psi_1}$ considering both tree-level and loop-level contributions shown by red solid line. }
    \label{fig:1}
\end{figure}

For details see the Ref.\cite{Hisano:2011cs,Essig:2007az,Hisano:2010fy,Amintaheri:2022coc}.
In Fig.\ref{fig:1} we show the variation of $\sigma_{\psi_1 N}$ with $M_{\psi_1}$ for loop level interactions and tree level interactions for different mixing angle 
$\alpha$.
So for $\sin \alpha=10^{-4}$, the direct detection cross-section is loop dominated.
%%%%%%%%%%%%%%%%%%%%%%%%%%%%%%%%%%%%%%%%%%5

\bibliography{ref}

\begin{thebibliography}{101}
\expandafter\ifx\csname natexlab\endcsname\relax\def\natexlab#1{#1}\fi
\expandafter\ifx\csname bibnamefont\endcsname\relax
  \def\bibnamefont#1{#1}\fi
\expandafter\ifx\csname bibfnamefont\endcsname\relax
  \def\bibfnamefont#1{#1}\fi
\expandafter\ifx\csname citenamefont\endcsname\relax
  \def\citenamefont#1{#1}\fi
\expandafter\ifx\csname url\endcsname\relax
  \def\url#1{\texttt{#1}}\fi
\expandafter\ifx\csname urlprefix\endcsname\relax\def\urlprefix{URL }\fi
\providecommand{\bibinfo}[2]{#2}
\providecommand{\eprint}[2][]{\url{#2}}

\bibitem[{\citenamefont{Zwicky}(1933)}]{Zwicky:1933gu}
\bibinfo{author}{\bibfnamefont{F.}~\bibnamefont{Zwicky}},
  \bibinfo{journal}{Helv. Phys. Acta} \textbf{\bibinfo{volume}{6}},
  \bibinfo{pages}{110} (\bibinfo{year}{1933}).

\bibitem[{\citenamefont{Rubin and Ford}(1970)}]{Rubin:1970zza}
\bibinfo{author}{\bibfnamefont{V.~C.} \bibnamefont{Rubin}} \bibnamefont{and}
  \bibinfo{author}{\bibfnamefont{W.~K.} \bibnamefont{Ford},
  \bibfnamefont{Jr.}}, \bibinfo{journal}{Astrophys. J.}
  \textbf{\bibinfo{volume}{159}}, \bibinfo{pages}{379} (\bibinfo{year}{1970}).

\bibitem[{\citenamefont{Clowe et~al.}(2006)\citenamefont{Clowe, Bradac,
  Gonzalez, Markevitch, Randall, Jones, and Zaritsky}}]{Clowe:2006eq}
\bibinfo{author}{\bibfnamefont{D.}~\bibnamefont{Clowe}},
  \bibinfo{author}{\bibfnamefont{M.}~\bibnamefont{Bradac}},
  \bibinfo{author}{\bibfnamefont{A.~H.} \bibnamefont{Gonzalez}},
  \bibinfo{author}{\bibfnamefont{M.}~\bibnamefont{Markevitch}},
  \bibinfo{author}{\bibfnamefont{S.~W.} \bibnamefont{Randall}},
  \bibinfo{author}{\bibfnamefont{C.}~\bibnamefont{Jones}}, \bibnamefont{and}
  \bibinfo{author}{\bibfnamefont{D.}~\bibnamefont{Zaritsky}},
  \bibinfo{journal}{Astrophys. J. Lett.} \textbf{\bibinfo{volume}{648}},
  \bibinfo{pages}{L109} (\bibinfo{year}{2006}), \eprint{astro-ph/0608407}.

\bibitem[{\citenamefont{Aghanim et~al.}(2020)}]{Planck:2018vyg}
\bibinfo{author}{\bibfnamefont{N.}~\bibnamefont{Aghanim}} \bibnamefont{et~al.}
  (\bibinfo{collaboration}{Planck}), \bibinfo{journal}{Astron. Astrophys.}
  \textbf{\bibinfo{volume}{641}}, \bibinfo{pages}{A6} (\bibinfo{year}{2020}),
  \bibinfo{note}{[Erratum: Astron.Astrophys. 652, C4 (2021)]},
  \eprint{1807.06209}.

\bibitem[{\citenamefont{Abe et~al.}(2011)}]{T2K:2011ypd}
\bibinfo{author}{\bibfnamefont{K.}~\bibnamefont{Abe}} \bibnamefont{et~al.}
  (\bibinfo{collaboration}{T2K}), \bibinfo{journal}{Phys. Rev. Lett.}
  \textbf{\bibinfo{volume}{107}}, \bibinfo{pages}{041801}
  (\bibinfo{year}{2011}), \eprint{1106.2822}.

\bibitem[{\citenamefont{Abe et~al.}(2012)}]{DoubleChooz:2011ymz}
\bibinfo{author}{\bibfnamefont{Y.}~\bibnamefont{Abe}} \bibnamefont{et~al.}
  (\bibinfo{collaboration}{Double Chooz}), \bibinfo{journal}{Phys. Rev. Lett.}
  \textbf{\bibinfo{volume}{108}}, \bibinfo{pages}{131801}
  (\bibinfo{year}{2012}), \eprint{1112.6353}.

\bibitem[{\citenamefont{An et~al.}(2012)}]{DayaBay:2012fng}
\bibinfo{author}{\bibfnamefont{F.~P.} \bibnamefont{An}} \bibnamefont{et~al.}
  (\bibinfo{collaboration}{Daya Bay}), \bibinfo{journal}{Phys. Rev. Lett.}
  \textbf{\bibinfo{volume}{108}}, \bibinfo{pages}{171803}
  (\bibinfo{year}{2012}), \eprint{1203.1669}.

\bibitem[{\citenamefont{Ahn et~al.}(2012)}]{RENO:2012mkc}
\bibinfo{author}{\bibfnamefont{J.~K.} \bibnamefont{Ahn}} \bibnamefont{et~al.}
  (\bibinfo{collaboration}{RENO}), \bibinfo{journal}{Phys. Rev. Lett.}
  \textbf{\bibinfo{volume}{108}}, \bibinfo{pages}{191802}
  (\bibinfo{year}{2012}), \eprint{1204.0626}.

\bibitem[{\citenamefont{Adamson et~al.}(2013)}]{MINOS:2013xrl}
\bibinfo{author}{\bibfnamefont{P.}~\bibnamefont{Adamson}} \bibnamefont{et~al.}
  (\bibinfo{collaboration}{MINOS}), \bibinfo{journal}{Phys. Rev. Lett.}
  \textbf{\bibinfo{volume}{110}}, \bibinfo{pages}{171801}
  (\bibinfo{year}{2013}), \eprint{1301.4581}.

\bibitem[{\citenamefont{Zyla et~al.}(2020)}]{ParticleDataGroup:2020ssz}
\bibinfo{author}{\bibfnamefont{P.~A.} \bibnamefont{Zyla}} \bibnamefont{et~al.}
  (\bibinfo{collaboration}{Particle Data Group}), \bibinfo{journal}{PTEP}
  \textbf{\bibinfo{volume}{2020}}, \bibinfo{pages}{083C01}
  (\bibinfo{year}{2020}).

\bibitem[{\citenamefont{Minkowski}(1977)}]{Minkowski:1977sc}
\bibinfo{author}{\bibfnamefont{P.}~\bibnamefont{Minkowski}},
  \bibinfo{journal}{Phys. Lett. B} \textbf{\bibinfo{volume}{67}},
  \bibinfo{pages}{421} (\bibinfo{year}{1977}).

\bibitem[{\citenamefont{Mohapatra and Senjanovic}(1980)}]{Mohapatra:1979ia}
\bibinfo{author}{\bibfnamefont{R.~N.} \bibnamefont{Mohapatra}}
  \bibnamefont{and}
  \bibinfo{author}{\bibfnamefont{G.}~\bibnamefont{Senjanovic}},
  \bibinfo{journal}{Phys. Rev. Lett.} \textbf{\bibinfo{volume}{44}},
  \bibinfo{pages}{912} (\bibinfo{year}{1980}).

\bibitem[{\citenamefont{Schechter and Valle}(1980)}]{Schechter:1980gr}
\bibinfo{author}{\bibfnamefont{J.}~\bibnamefont{Schechter}} \bibnamefont{and}
  \bibinfo{author}{\bibfnamefont{J.~W.~F.} \bibnamefont{Valle}},
  \bibinfo{journal}{Phys. Rev. D} \textbf{\bibinfo{volume}{22}},
  \bibinfo{pages}{2227} (\bibinfo{year}{1980}).

\bibitem[{\citenamefont{Gell-Mann et~al.}(1979)\citenamefont{Gell-Mann, Ramond,
  and Slansky}}]{Gell-Mann:1979vob}
\bibinfo{author}{\bibfnamefont{M.}~\bibnamefont{Gell-Mann}},
  \bibinfo{author}{\bibfnamefont{P.}~\bibnamefont{Ramond}}, \bibnamefont{and}
  \bibinfo{author}{\bibfnamefont{R.}~\bibnamefont{Slansky}},
  \bibinfo{journal}{Conf. Proc. C} \textbf{\bibinfo{volume}{790927}},
  \bibinfo{pages}{315} (\bibinfo{year}{1979}), \eprint{1306.4669}.

\bibitem[{\citenamefont{Mohapatra and Senjanovic}(1981)}]{Mohapatra:1980yp}
\bibinfo{author}{\bibfnamefont{R.~N.} \bibnamefont{Mohapatra}}
  \bibnamefont{and}
  \bibinfo{author}{\bibfnamefont{G.}~\bibnamefont{Senjanovic}},
  \bibinfo{journal}{Phys. Rev. D} \textbf{\bibinfo{volume}{23}},
  \bibinfo{pages}{165} (\bibinfo{year}{1981}).

\bibitem[{\citenamefont{Lazarides et~al.}(1981)\citenamefont{Lazarides, Shafi,
  and Wetterich}}]{Lazarides:1980nt}
\bibinfo{author}{\bibfnamefont{G.}~\bibnamefont{Lazarides}},
  \bibinfo{author}{\bibfnamefont{Q.}~\bibnamefont{Shafi}}, \bibnamefont{and}
  \bibinfo{author}{\bibfnamefont{C.}~\bibnamefont{Wetterich}},
  \bibinfo{journal}{Nucl. Phys. B} \textbf{\bibinfo{volume}{181}},
  \bibinfo{pages}{287} (\bibinfo{year}{1981}).

\bibitem[{\citenamefont{Wetterich}(1981)}]{Wetterich:1981bx}
\bibinfo{author}{\bibfnamefont{C.}~\bibnamefont{Wetterich}},
  \bibinfo{journal}{Nucl. Phys. B} \textbf{\bibinfo{volume}{187}},
  \bibinfo{pages}{343} (\bibinfo{year}{1981}).

\bibitem[{\citenamefont{Schechter and Valle}(1982)}]{Schechter:1981cv}
\bibinfo{author}{\bibfnamefont{J.}~\bibnamefont{Schechter}} \bibnamefont{and}
  \bibinfo{author}{\bibfnamefont{J.~W.~F.} \bibnamefont{Valle}},
  \bibinfo{journal}{Phys. Rev. D} \textbf{\bibinfo{volume}{25}},
  \bibinfo{pages}{774} (\bibinfo{year}{1982}).

\bibitem[{\citenamefont{Brahmachari and Mohapatra}(1998)}]{Brahmachari:1997cq}
\bibinfo{author}{\bibfnamefont{B.}~\bibnamefont{Brahmachari}} \bibnamefont{and}
  \bibinfo{author}{\bibfnamefont{R.~N.} \bibnamefont{Mohapatra}},
  \bibinfo{journal}{Phys. Rev. D} \textbf{\bibinfo{volume}{58}},
  \bibinfo{pages}{015001} (\bibinfo{year}{1998}), \eprint{hep-ph/9710371}.

\bibitem[{\citenamefont{Foot et~al.}(1989)\citenamefont{Foot, Lew, He, and
  Joshi}}]{Foot:1988aq}
\bibinfo{author}{\bibfnamefont{R.}~\bibnamefont{Foot}},
  \bibinfo{author}{\bibfnamefont{H.}~\bibnamefont{Lew}},
  \bibinfo{author}{\bibfnamefont{X.~G.} \bibnamefont{He}}, \bibnamefont{and}
  \bibinfo{author}{\bibfnamefont{G.~C.} \bibnamefont{Joshi}},
  \bibinfo{journal}{Z. Phys. C} \textbf{\bibinfo{volume}{44}},
  \bibinfo{pages}{441} (\bibinfo{year}{1989}).

\bibitem[{\citenamefont{Kolb and Turner}(1990)}]{Kolb:1990vq}
\bibinfo{author}{\bibfnamefont{E.~W.} \bibnamefont{Kolb}} \bibnamefont{and}
  \bibinfo{author}{\bibfnamefont{M.~S.} \bibnamefont{Turner}},
  \emph{\bibinfo{title}{{The Early Universe}}}, vol.~\bibinfo{volume}{69}
  (\bibinfo{year}{1990}), ISBN \bibinfo{isbn}{978-0-201-62674-2}.

\bibitem[{\citenamefont{Feng}(2010)}]{Feng:2010gw}
\bibinfo{author}{\bibfnamefont{J.~L.} \bibnamefont{Feng}},
  \bibinfo{journal}{Ann. Rev. Astron. Astrophys.}
  \textbf{\bibinfo{volume}{48}}, \bibinfo{pages}{495} (\bibinfo{year}{2010}),
  \eprint{1003.0904}.

\bibitem[{\citenamefont{Roszkowski et~al.}(2018)\citenamefont{Roszkowski,
  Sessolo, and Trojanowski}}]{Roszkowski:2017nbc}
\bibinfo{author}{\bibfnamefont{L.}~\bibnamefont{Roszkowski}},
  \bibinfo{author}{\bibfnamefont{E.~M.} \bibnamefont{Sessolo}},
  \bibnamefont{and}
  \bibinfo{author}{\bibfnamefont{S.}~\bibnamefont{Trojanowski}},
  \bibinfo{journal}{Rept. Prog. Phys.} \textbf{\bibinfo{volume}{81}},
  \bibinfo{pages}{066201} (\bibinfo{year}{2018}), \eprint{1707.06277}.

\bibitem[{\citenamefont{Schumann}(2019)}]{Schumann:2019eaa}
\bibinfo{author}{\bibfnamefont{M.}~\bibnamefont{Schumann}},
  \bibinfo{journal}{J. Phys. G} \textbf{\bibinfo{volume}{46}},
  \bibinfo{pages}{103003} (\bibinfo{year}{2019}), \eprint{1903.03026}.

\bibitem[{\citenamefont{Lin}(2019)}]{Lin:2019uvt}
\bibinfo{author}{\bibfnamefont{T.}~\bibnamefont{Lin}}, \bibinfo{journal}{PoS}
  \textbf{\bibinfo{volume}{333}}, \bibinfo{pages}{009} (\bibinfo{year}{2019}),
  \eprint{1904.07915}.

\bibitem[{\citenamefont{Arcadi et~al.}(2018)\citenamefont{Arcadi, Dutra, Ghosh,
  Lindner, Mambrini, Pierre, Profumo, and Queiroz}}]{Arcadi:2017kky}
\bibinfo{author}{\bibfnamefont{G.}~\bibnamefont{Arcadi}},
  \bibinfo{author}{\bibfnamefont{M.}~\bibnamefont{Dutra}},
  \bibinfo{author}{\bibfnamefont{P.}~\bibnamefont{Ghosh}},
  \bibinfo{author}{\bibfnamefont{M.}~\bibnamefont{Lindner}},
  \bibinfo{author}{\bibfnamefont{Y.}~\bibnamefont{Mambrini}},
  \bibinfo{author}{\bibfnamefont{M.}~\bibnamefont{Pierre}},
  \bibinfo{author}{\bibfnamefont{S.}~\bibnamefont{Profumo}}, \bibnamefont{and}
  \bibinfo{author}{\bibfnamefont{F.~S.} \bibnamefont{Queiroz}},
  \bibinfo{journal}{Eur. Phys. J. C} \textbf{\bibinfo{volume}{78}},
  \bibinfo{pages}{203} (\bibinfo{year}{2018}), \eprint{1703.07364}.

\bibitem[{\citenamefont{Tan et~al.}(2016)}]{PandaX-II:2016vec}
\bibinfo{author}{\bibfnamefont{A.}~\bibnamefont{Tan}} \bibnamefont{et~al.}
  (\bibinfo{collaboration}{PandaX-II}), \bibinfo{journal}{Phys. Rev. Lett.}
  \textbf{\bibinfo{volume}{117}}, \bibinfo{pages}{121303}
  (\bibinfo{year}{2016}), \eprint{1607.07400}.

\bibitem[{\citenamefont{Aprile et~al.}(2017)}]{XENON:2017vdw}
\bibinfo{author}{\bibfnamefont{E.}~\bibnamefont{Aprile}} \bibnamefont{et~al.}
  (\bibinfo{collaboration}{XENON}), \bibinfo{journal}{Phys. Rev. Lett.}
  \textbf{\bibinfo{volume}{119}}, \bibinfo{pages}{181301}
  (\bibinfo{year}{2017}), \eprint{1705.06655}.

\bibitem[{\citenamefont{Akerib et~al.}(2017)}]{LUX:2016ggv}
\bibinfo{author}{\bibfnamefont{D.~S.} \bibnamefont{Akerib}}
  \bibnamefont{et~al.} (\bibinfo{collaboration}{LUX}), \bibinfo{journal}{Phys.
  Rev. Lett.} \textbf{\bibinfo{volume}{118}}, \bibinfo{pages}{021303}
  (\bibinfo{year}{2017}), \eprint{1608.07648}.

\bibitem[{\citenamefont{Amole et~al.}(2019)}]{PICO:2019vsc}
\bibinfo{author}{\bibfnamefont{C.}~\bibnamefont{Amole}} \bibnamefont{et~al.}
  (\bibinfo{collaboration}{PICO}), \bibinfo{journal}{Phys. Rev. D}
  \textbf{\bibinfo{volume}{100}}, \bibinfo{pages}{022001}
  (\bibinfo{year}{2019}), \eprint{1902.04031}.

\bibitem[{\citenamefont{Aguilar et~al.}(2013)}]{AMS:2013fma}
\bibinfo{author}{\bibfnamefont{M.}~\bibnamefont{Aguilar}} \bibnamefont{et~al.}
  (\bibinfo{collaboration}{AMS}), \bibinfo{journal}{Phys. Rev. Lett.}
  \textbf{\bibinfo{volume}{110}}, \bibinfo{pages}{141102}
  (\bibinfo{year}{2013}).

\bibitem[{\citenamefont{Buckley et~al.}(2013)}]{Buckley:2013bha}
\bibinfo{author}{\bibfnamefont{J.}~\bibnamefont{Buckley}} \bibnamefont{et~al.}
  (\bibinfo{year}{2013}), \eprint{1310.7040}.

\bibitem[{\citenamefont{Gaskins}(2016)}]{Gaskins:2016cha}
\bibinfo{author}{\bibfnamefont{J.~M.} \bibnamefont{Gaskins}},
  \bibinfo{journal}{Contemp. Phys.} \textbf{\bibinfo{volume}{57}},
  \bibinfo{pages}{496} (\bibinfo{year}{2016}), \eprint{1604.00014}.

\bibitem[{\citenamefont{Albert et~al.}(2017)}]{Fermi-LAT:2016uux}
\bibinfo{author}{\bibfnamefont{A.}~\bibnamefont{Albert}} \bibnamefont{et~al.}
  (\bibinfo{collaboration}{Fermi-LAT, DES}), \bibinfo{journal}{Astrophys. J.}
  \textbf{\bibinfo{volume}{834}}, \bibinfo{pages}{110} (\bibinfo{year}{2017}),
  \eprint{1611.03184}.

\bibitem[{\citenamefont{Ahnen et~al.}(2016)}]{MAGIC:2016xys}
\bibinfo{author}{\bibfnamefont{M.~L.} \bibnamefont{Ahnen}} \bibnamefont{et~al.}
  (\bibinfo{collaboration}{MAGIC, Fermi-LAT}), \bibinfo{journal}{JCAP}
  \textbf{\bibinfo{volume}{02}}, \bibinfo{pages}{039} (\bibinfo{year}{2016}),
  \eprint{1601.06590}.

\bibitem[{\citenamefont{Bringmann and Weniger}(2012)}]{Bringmann:2012ez}
\bibinfo{author}{\bibfnamefont{T.}~\bibnamefont{Bringmann}} \bibnamefont{and}
  \bibinfo{author}{\bibfnamefont{C.}~\bibnamefont{Weniger}},
  \bibinfo{journal}{Phys. Dark Univ.} \textbf{\bibinfo{volume}{1}},
  \bibinfo{pages}{194} (\bibinfo{year}{2012}), \eprint{1208.5481}.

\bibitem[{\citenamefont{Cirelli}(2016)}]{Cirelli:2015gux}
\bibinfo{author}{\bibfnamefont{M.}~\bibnamefont{Cirelli}},
  \bibinfo{journal}{PoS} \textbf{\bibinfo{volume}{ICRC2015}},
  \bibinfo{pages}{014} (\bibinfo{year}{2016}), \eprint{1511.02031}.

\bibitem[{\citenamefont{Kahlhoefer}(2017)}]{Kahlhoefer:2017dnp}
\bibinfo{author}{\bibfnamefont{F.}~\bibnamefont{Kahlhoefer}},
  \bibinfo{journal}{Int. J. Mod. Phys. A} \textbf{\bibinfo{volume}{32}},
  \bibinfo{pages}{1730006} (\bibinfo{year}{2017}), \eprint{1702.02430}.

\bibitem[{\citenamefont{Boveia and Doglioni}(2018)}]{Boveia:2018yeb}
\bibinfo{author}{\bibfnamefont{A.}~\bibnamefont{Boveia}} \bibnamefont{and}
  \bibinfo{author}{\bibfnamefont{C.}~\bibnamefont{Doglioni}},
  \bibinfo{journal}{Ann. Rev. Nucl. Part. Sci.} \textbf{\bibinfo{volume}{68}},
  \bibinfo{pages}{429} (\bibinfo{year}{2018}), \eprint{1810.12238}.

\bibitem[{\citenamefont{Hall et~al.}(2010)\citenamefont{Hall, Jedamzik,
  March-Russell, and West}}]{Hall:2009bx}
\bibinfo{author}{\bibfnamefont{L.~J.} \bibnamefont{Hall}},
  \bibinfo{author}{\bibfnamefont{K.}~\bibnamefont{Jedamzik}},
  \bibinfo{author}{\bibfnamefont{J.}~\bibnamefont{March-Russell}},
  \bibnamefont{and} \bibinfo{author}{\bibfnamefont{S.~M.} \bibnamefont{West}},
  \bibinfo{journal}{JHEP} \textbf{\bibinfo{volume}{03}}, \bibinfo{pages}{080}
  (\bibinfo{year}{2010}), \eprint{0911.1120}.

\bibitem[{\citenamefont{K\"onig et~al.}(2016)\citenamefont{K\"onig, Merle, and
  Totzauer}}]{Konig:2016dzg}
\bibinfo{author}{\bibfnamefont{J.}~\bibnamefont{K\"onig}},
  \bibinfo{author}{\bibfnamefont{A.}~\bibnamefont{Merle}}, \bibnamefont{and}
  \bibinfo{author}{\bibfnamefont{M.}~\bibnamefont{Totzauer}},
  \bibinfo{journal}{JCAP} \textbf{\bibinfo{volume}{11}}, \bibinfo{pages}{038}
  (\bibinfo{year}{2016}), \eprint{1609.01289}.

\bibitem[{\citenamefont{Biswas and Gupta}(2016)}]{Biswas:2016bfo}
\bibinfo{author}{\bibfnamefont{A.}~\bibnamefont{Biswas}} \bibnamefont{and}
  \bibinfo{author}{\bibfnamefont{A.}~\bibnamefont{Gupta}},
  \bibinfo{journal}{JCAP} \textbf{\bibinfo{volume}{09}}, \bibinfo{pages}{044}
  (\bibinfo{year}{2016}), \bibinfo{note}{[Addendum: JCAP 05, A01 (2017)]},
  \eprint{1607.01469}.

\bibitem[{\citenamefont{Bernal et~al.}(2017)\citenamefont{Bernal, Heikinheimo,
  Tenkanen, Tuominen, and Vaskonen}}]{Bernal:2017kxu}
\bibinfo{author}{\bibfnamefont{N.}~\bibnamefont{Bernal}},
  \bibinfo{author}{\bibfnamefont{M.}~\bibnamefont{Heikinheimo}},
  \bibinfo{author}{\bibfnamefont{T.}~\bibnamefont{Tenkanen}},
  \bibinfo{author}{\bibfnamefont{K.}~\bibnamefont{Tuominen}}, \bibnamefont{and}
  \bibinfo{author}{\bibfnamefont{V.}~\bibnamefont{Vaskonen}},
  \bibinfo{journal}{Int. J. Mod. Phys. A} \textbf{\bibinfo{volume}{32}},
  \bibinfo{pages}{1730023} (\bibinfo{year}{2017}), \eprint{1706.07442}.

\bibitem[{\citenamefont{Borah et~al.}(2018)\citenamefont{Borah, Karmakar, and
  Nanda}}]{Borah:2018gjk}
\bibinfo{author}{\bibfnamefont{D.}~\bibnamefont{Borah}},
  \bibinfo{author}{\bibfnamefont{B.}~\bibnamefont{Karmakar}}, \bibnamefont{and}
  \bibinfo{author}{\bibfnamefont{D.}~\bibnamefont{Nanda}},
  \bibinfo{journal}{JCAP} \textbf{\bibinfo{volume}{07}}, \bibinfo{pages}{039}
  (\bibinfo{year}{2018}), \eprint{1805.11115}.

\bibitem[{\citenamefont{Ghosh et~al.}(2023)\citenamefont{Ghosh, Ghosh, and
  Jeesun}}]{Ghosh:2023ocl}
\bibinfo{author}{\bibfnamefont{D.~K.} \bibnamefont{Ghosh}},
  \bibinfo{author}{\bibfnamefont{P.}~\bibnamefont{Ghosh}}, \bibnamefont{and}
  \bibinfo{author}{\bibfnamefont{S.}~\bibnamefont{Jeesun}}
  (\bibinfo{year}{2023}), \eprint{2301.13754}.

\bibitem[{\citenamefont{Hochberg et~al.}(2014)\citenamefont{Hochberg, Kuflik,
  Volansky, and Wacker}}]{Hochberg:2014dra}
\bibinfo{author}{\bibfnamefont{Y.}~\bibnamefont{Hochberg}},
  \bibinfo{author}{\bibfnamefont{E.}~\bibnamefont{Kuflik}},
  \bibinfo{author}{\bibfnamefont{T.}~\bibnamefont{Volansky}}, \bibnamefont{and}
  \bibinfo{author}{\bibfnamefont{J.~G.} \bibnamefont{Wacker}},
  \bibinfo{journal}{Phys. Rev. Lett.} \textbf{\bibinfo{volume}{113}},
  \bibinfo{pages}{171301} (\bibinfo{year}{2014}), \eprint{1402.5143}.

\bibitem[{\citenamefont{D'Eramo et~al.}(2017)\citenamefont{D'Eramo, Fernandez,
  and Profumo}}]{DEramo:2017gpl}
\bibinfo{author}{\bibfnamefont{F.}~\bibnamefont{D'Eramo}},
  \bibinfo{author}{\bibfnamefont{N.}~\bibnamefont{Fernandez}},
  \bibnamefont{and} \bibinfo{author}{\bibfnamefont{S.}~\bibnamefont{Profumo}},
  \bibinfo{journal}{JCAP} \textbf{\bibinfo{volume}{05}}, \bibinfo{pages}{012}
  (\bibinfo{year}{2017}), \eprint{1703.04793}.

\bibitem[{\citenamefont{Medina}(2017)}]{Medina:2014bga}
\bibinfo{author}{\bibfnamefont{A.~D.} \bibnamefont{Medina}},
  \bibinfo{journal}{Phys. Lett. B} \textbf{\bibinfo{volume}{770}},
  \bibinfo{pages}{161} (\bibinfo{year}{2017}), \eprint{1409.2560}.

\bibitem[{\citenamefont{Puetter et~al.}(2022)\citenamefont{Puetter, Ruderman,
  Salvioni, and Shakya}}]{Puetter:2022ucx}
\bibinfo{author}{\bibfnamefont{L.}~\bibnamefont{Puetter}},
  \bibinfo{author}{\bibfnamefont{J.~T.} \bibnamefont{Ruderman}},
  \bibinfo{author}{\bibfnamefont{E.}~\bibnamefont{Salvioni}}, \bibnamefont{and}
  \bibinfo{author}{\bibfnamefont{B.}~\bibnamefont{Shakya}}
  (\bibinfo{year}{2022}), \eprint{2208.08453}.

\bibitem[{\citenamefont{Frumkin et~al.}(2023)\citenamefont{Frumkin, Hochberg,
  Kuflik, and Murayama}}]{Frumkin:2021zng}
\bibinfo{author}{\bibfnamefont{R.}~\bibnamefont{Frumkin}},
  \bibinfo{author}{\bibfnamefont{Y.}~\bibnamefont{Hochberg}},
  \bibinfo{author}{\bibfnamefont{E.}~\bibnamefont{Kuflik}}, \bibnamefont{and}
  \bibinfo{author}{\bibfnamefont{H.}~\bibnamefont{Murayama}},
  \bibinfo{journal}{Phys. Rev. Lett.} \textbf{\bibinfo{volume}{130}},
  \bibinfo{pages}{121001} (\bibinfo{year}{2023}), \eprint{2111.14857}.

\bibitem[{\citenamefont{Fairbairn and Zupan}(2009)}]{Fairbairn:2008fb}
\bibinfo{author}{\bibfnamefont{M.}~\bibnamefont{Fairbairn}} \bibnamefont{and}
  \bibinfo{author}{\bibfnamefont{J.}~\bibnamefont{Zupan}},
  \bibinfo{journal}{JCAP} \textbf{\bibinfo{volume}{07}}, \bibinfo{pages}{001}
  (\bibinfo{year}{2009}), \eprint{0810.4147}.

\bibitem[{\citenamefont{D\'\i{}az~S\'aez and
  Contreras}(2023)}]{DiazSaez:2023wli}
\bibinfo{author}{\bibfnamefont{B.}~\bibnamefont{D\'\i{}az~S\'aez}}
  \bibnamefont{and} \bibinfo{author}{\bibfnamefont{P.~E.}
  \bibnamefont{Contreras}} (\bibinfo{year}{2023}), \eprint{2307.07760}.

\bibitem[{\citenamefont{Borah and Gupta}(2017)}]{Borah:2017dfn}
\bibinfo{author}{\bibfnamefont{D.}~\bibnamefont{Borah}} \bibnamefont{and}
  \bibinfo{author}{\bibfnamefont{A.}~\bibnamefont{Gupta}},
  \bibinfo{journal}{Phys. Rev. D} \textbf{\bibinfo{volume}{96}},
  \bibinfo{pages}{115012} (\bibinfo{year}{2017}), \eprint{1706.05034}.

\bibitem[{\citenamefont{Biswas et~al.}(2018)\citenamefont{Biswas, Borah, and
  Nanda}}]{Biswas:2018ybc}
\bibinfo{author}{\bibfnamefont{A.}~\bibnamefont{Biswas}},
  \bibinfo{author}{\bibfnamefont{D.}~\bibnamefont{Borah}}, \bibnamefont{and}
  \bibinfo{author}{\bibfnamefont{D.}~\bibnamefont{Nanda}},
  \bibinfo{journal}{JCAP} \textbf{\bibinfo{volume}{09}}, \bibinfo{pages}{014}
  (\bibinfo{year}{2018}), \eprint{1806.01876}.

\bibitem[{\citenamefont{Ma}(2006)}]{Ma:2006km}
\bibinfo{author}{\bibfnamefont{E.}~\bibnamefont{Ma}}, \bibinfo{journal}{Phys.
  Rev. D} \textbf{\bibinfo{volume}{73}}, \bibinfo{pages}{077301}
  (\bibinfo{year}{2006}), \eprint{hep-ph/0601225}.

\bibitem[{\citenamefont{Ma and Suematsu}(2009)}]{Ma:2008cu}
\bibinfo{author}{\bibfnamefont{E.}~\bibnamefont{Ma}} \bibnamefont{and}
  \bibinfo{author}{\bibfnamefont{D.}~\bibnamefont{Suematsu}},
  \bibinfo{journal}{Mod. Phys. Lett. A} \textbf{\bibinfo{volume}{24}},
  \bibinfo{pages}{583} (\bibinfo{year}{2009}), \eprint{0809.0942}.

\bibitem[{\citenamefont{Arkani-Hamed et~al.}(2005)\citenamefont{Arkani-Hamed,
  Dimopoulos, and Kachru}}]{Arkani-Hamed:2005zuc}
\bibinfo{author}{\bibfnamefont{N.}~\bibnamefont{Arkani-Hamed}},
  \bibinfo{author}{\bibfnamefont{S.}~\bibnamefont{Dimopoulos}},
  \bibnamefont{and} \bibinfo{author}{\bibfnamefont{S.}~\bibnamefont{Kachru}}
  (\bibinfo{year}{2005}), \eprint{hep-th/0501082}.

\bibitem[{\citenamefont{D'Eramo}(2007)}]{DEramo:2007anh}
\bibinfo{author}{\bibfnamefont{F.}~\bibnamefont{D'Eramo}},
  \bibinfo{journal}{Phys. Rev. D} \textbf{\bibinfo{volume}{76}},
  \bibinfo{pages}{083522} (\bibinfo{year}{2007}), \eprint{0705.4493}.

\bibitem[{\citenamefont{Bhattacharya
  et~al.}(2019{\natexlab{a}})\citenamefont{Bhattacharya, Ghosh, Sahoo, and
  Sahu}}]{Bhattacharya:2018fus}
\bibinfo{author}{\bibfnamefont{S.}~\bibnamefont{Bhattacharya}},
  \bibinfo{author}{\bibfnamefont{P.}~\bibnamefont{Ghosh}},
  \bibinfo{author}{\bibfnamefont{N.}~\bibnamefont{Sahoo}}, \bibnamefont{and}
  \bibinfo{author}{\bibfnamefont{N.}~\bibnamefont{Sahu}},
  \bibinfo{journal}{Front. in Phys.} \textbf{\bibinfo{volume}{7}},
  \bibinfo{pages}{80} (\bibinfo{year}{2019}{\natexlab{a}}),
  \eprint{1812.06505}.

\bibitem[{\citenamefont{Mahbubani and Senatore}(2006)}]{Mahbubani:2005pt}
\bibinfo{author}{\bibfnamefont{R.}~\bibnamefont{Mahbubani}} \bibnamefont{and}
  \bibinfo{author}{\bibfnamefont{L.}~\bibnamefont{Senatore}},
  \bibinfo{journal}{Phys. Rev. D} \textbf{\bibinfo{volume}{73}},
  \bibinfo{pages}{043510} (\bibinfo{year}{2006}), \eprint{hep-ph/0510064}.

\bibitem[{\citenamefont{Bhattacharya et~al.}(2016)\citenamefont{Bhattacharya,
  Sahoo, and Sahu}}]{Bhattacharya:2015qpa}
\bibinfo{author}{\bibfnamefont{S.}~\bibnamefont{Bhattacharya}},
  \bibinfo{author}{\bibfnamefont{N.}~\bibnamefont{Sahoo}}, \bibnamefont{and}
  \bibinfo{author}{\bibfnamefont{N.}~\bibnamefont{Sahu}},
  \bibinfo{journal}{Phys. Rev. D} \textbf{\bibinfo{volume}{93}},
  \bibinfo{pages}{115040} (\bibinfo{year}{2016}), \eprint{1510.02760}.

\bibitem[{\citenamefont{Bhattacharya
  et~al.}(2017{\natexlab{a}})\citenamefont{Bhattacharya, Sahoo, and
  Sahu}}]{Bhattacharya:2017sml}
\bibinfo{author}{\bibfnamefont{S.}~\bibnamefont{Bhattacharya}},
  \bibinfo{author}{\bibfnamefont{N.}~\bibnamefont{Sahoo}}, \bibnamefont{and}
  \bibinfo{author}{\bibfnamefont{N.}~\bibnamefont{Sahu}},
  \bibinfo{journal}{Phys. Rev. D} \textbf{\bibinfo{volume}{96}},
  \bibinfo{pages}{035010} (\bibinfo{year}{2017}{\natexlab{a}}),
  \eprint{1704.03417}.

\bibitem[{\citenamefont{Barman et~al.}(2019{\natexlab{a}})\citenamefont{Barman,
  Borah, Ghosh, and Saha}}]{Barman:2019aku}
\bibinfo{author}{\bibfnamefont{B.}~\bibnamefont{Barman}},
  \bibinfo{author}{\bibfnamefont{D.}~\bibnamefont{Borah}},
  \bibinfo{author}{\bibfnamefont{P.}~\bibnamefont{Ghosh}}, \bibnamefont{and}
  \bibinfo{author}{\bibfnamefont{A.~K.} \bibnamefont{Saha}},
  \bibinfo{journal}{JHEP} \textbf{\bibinfo{volume}{10}}, \bibinfo{pages}{275}
  (\bibinfo{year}{2019}{\natexlab{a}}), \eprint{1907.10071}.

\bibitem[{\citenamefont{Ghosh et~al.}(2022{\natexlab{a}})\citenamefont{Ghosh,
  Konar, Saha, and Show}}]{Ghosh:2021wrk}
\bibinfo{author}{\bibfnamefont{P.}~\bibnamefont{Ghosh}},
  \bibinfo{author}{\bibfnamefont{P.}~\bibnamefont{Konar}},
  \bibinfo{author}{\bibfnamefont{A.~K.} \bibnamefont{Saha}}, \bibnamefont{and}
  \bibinfo{author}{\bibfnamefont{S.}~\bibnamefont{Show}},
  \bibinfo{journal}{JCAP} \textbf{\bibinfo{volume}{10}}, \bibinfo{pages}{017}
  (\bibinfo{year}{2022}{\natexlab{a}}), \eprint{2112.09057}.

\bibitem[{\citenamefont{Dey et~al.}(2022)\citenamefont{Dey, Ghosh, and
  Rai}}]{Dey:2022whc}
\bibinfo{author}{\bibfnamefont{S.}~\bibnamefont{Dey}},
  \bibinfo{author}{\bibfnamefont{P.}~\bibnamefont{Ghosh}}, \bibnamefont{and}
  \bibinfo{author}{\bibfnamefont{S.~K.} \bibnamefont{Rai}},
  \bibinfo{journal}{Eur. Phys. J. C} \textbf{\bibinfo{volume}{82}},
  \bibinfo{pages}{876} (\bibinfo{year}{2022}), \eprint{2202.11638}.

\bibitem[{\citenamefont{Bhattacharya
  et~al.}(2022{\natexlab{a}})\citenamefont{Bhattacharya, Jahedi, and
  Wudka}}]{Bhattacharya:2021ltd}
\bibinfo{author}{\bibfnamefont{S.}~\bibnamefont{Bhattacharya}},
  \bibinfo{author}{\bibfnamefont{S.}~\bibnamefont{Jahedi}}, \bibnamefont{and}
  \bibinfo{author}{\bibfnamefont{J.}~\bibnamefont{Wudka}},
  \bibinfo{journal}{JHEP} \textbf{\bibinfo{volume}{05}}, \bibinfo{pages}{009}
  (\bibinfo{year}{2022}{\natexlab{a}}), \eprint{2106.02846}.

\bibitem[{\citenamefont{Konar et~al.}(2020)\citenamefont{Konar, Mukherjee,
  Saha, and Show}}]{Konar:2020wvl}
\bibinfo{author}{\bibfnamefont{P.}~\bibnamefont{Konar}},
  \bibinfo{author}{\bibfnamefont{A.}~\bibnamefont{Mukherjee}},
  \bibinfo{author}{\bibfnamefont{A.~K.} \bibnamefont{Saha}}, \bibnamefont{and}
  \bibinfo{author}{\bibfnamefont{S.}~\bibnamefont{Show}},
  \bibinfo{journal}{Phys. Rev. D} \textbf{\bibinfo{volume}{102}},
  \bibinfo{pages}{015024} (\bibinfo{year}{2020}), \eprint{2001.11325}.

\bibitem[{\citenamefont{Konar et~al.}(2021)\citenamefont{Konar, Mukherjee,
  Saha, and Show}}]{Konar:2020vuu}
\bibinfo{author}{\bibfnamefont{P.}~\bibnamefont{Konar}},
  \bibinfo{author}{\bibfnamefont{A.}~\bibnamefont{Mukherjee}},
  \bibinfo{author}{\bibfnamefont{A.~K.} \bibnamefont{Saha}}, \bibnamefont{and}
  \bibinfo{author}{\bibfnamefont{S.}~\bibnamefont{Show}},
  \bibinfo{journal}{JHEP} \textbf{\bibinfo{volume}{03}}, \bibinfo{pages}{044}
  (\bibinfo{year}{2021}), \eprint{2007.15608}.

\bibitem[{\citenamefont{Bhattacharya
  et~al.}(2022{\natexlab{b}})\citenamefont{Bhattacharya, Ghosh, Lahiri, and
  Mukhopadhyaya}}]{Bhattacharya:2022wtr}
\bibinfo{author}{\bibfnamefont{S.}~\bibnamefont{Bhattacharya}},
  \bibinfo{author}{\bibfnamefont{P.}~\bibnamefont{Ghosh}},
  \bibinfo{author}{\bibfnamefont{J.}~\bibnamefont{Lahiri}}, \bibnamefont{and}
  \bibinfo{author}{\bibfnamefont{B.}~\bibnamefont{Mukhopadhyaya}},
  \bibinfo{journal}{JHEP} \textbf{\bibinfo{volume}{12}}, \bibinfo{pages}{049}
  (\bibinfo{year}{2022}{\natexlab{b}}), \eprint{2202.12097}.

\bibitem[{\citenamefont{Nollett and Steigman}(2015)}]{Nollett:2014lwa}
\bibinfo{author}{\bibfnamefont{K.~M.} \bibnamefont{Nollett}} \bibnamefont{and}
  \bibinfo{author}{\bibfnamefont{G.}~\bibnamefont{Steigman}},
  \bibinfo{journal}{Phys. Rev. D} \textbf{\bibinfo{volume}{91}},
  \bibinfo{pages}{083505} (\bibinfo{year}{2015}), \eprint{1411.6005}.

\bibitem[{\citenamefont{Mahanta and Borah}(2019)}]{Mahanta:2019gfe}
\bibinfo{author}{\bibfnamefont{D.}~\bibnamefont{Mahanta}} \bibnamefont{and}
  \bibinfo{author}{\bibfnamefont{D.}~\bibnamefont{Borah}},
  \bibinfo{journal}{JCAP} \textbf{\bibinfo{volume}{11}}, \bibinfo{pages}{021}
  (\bibinfo{year}{2019}), \eprint{1906.03577}.

\bibitem[{\citenamefont{Bonilla et~al.}(2020)\citenamefont{Bonilla, de~la Vega,
  Lamprea, Lineros, and Peinado}}]{Bonilla:2019ipe}
\bibinfo{author}{\bibfnamefont{C.}~\bibnamefont{Bonilla}},
  \bibinfo{author}{\bibfnamefont{L.~M.~G.} \bibnamefont{de~la Vega}},
  \bibinfo{author}{\bibfnamefont{J.~M.} \bibnamefont{Lamprea}},
  \bibinfo{author}{\bibfnamefont{R.~A.} \bibnamefont{Lineros}},
  \bibnamefont{and} \bibinfo{author}{\bibfnamefont{E.}~\bibnamefont{Peinado}},
  \bibinfo{journal}{New J. Phys.} \textbf{\bibinfo{volume}{22}},
  \bibinfo{pages}{033009} (\bibinfo{year}{2020}), \eprint{1908.04276}.

\bibitem[{\citenamefont{Drees et~al.}(2006)\citenamefont{Drees, Iminniyaz, and
  Kakizaki}}]{Drees:2006vh}
\bibinfo{author}{\bibfnamefont{M.}~\bibnamefont{Drees}},
  \bibinfo{author}{\bibfnamefont{H.}~\bibnamefont{Iminniyaz}},
  \bibnamefont{and} \bibinfo{author}{\bibfnamefont{M.}~\bibnamefont{Kakizaki}},
  \bibinfo{journal}{Phys. Rev. D} \textbf{\bibinfo{volume}{73}},
  \bibinfo{pages}{123502} (\bibinfo{year}{2006}), \eprint{hep-ph/0603165}.

\bibitem[{\citenamefont{Giudice et~al.}(2001)\citenamefont{Giudice, Kolb, and
  Riotto}}]{Giudice:2000ex}
\bibinfo{author}{\bibfnamefont{G.~F.} \bibnamefont{Giudice}},
  \bibinfo{author}{\bibfnamefont{E.~W.} \bibnamefont{Kolb}}, \bibnamefont{and}
  \bibinfo{author}{\bibfnamefont{A.}~\bibnamefont{Riotto}},
  \bibinfo{journal}{Phys. Rev. D} \textbf{\bibinfo{volume}{64}},
  \bibinfo{pages}{023508} (\bibinfo{year}{2001}), \eprint{hep-ph/0005123}.

\bibitem[{\citenamefont{Thomas and Wells}(1998)}]{Thomas:1998wy}
\bibinfo{author}{\bibfnamefont{S.~D.} \bibnamefont{Thomas}} \bibnamefont{and}
  \bibinfo{author}{\bibfnamefont{J.~D.} \bibnamefont{Wells}},
  \bibinfo{journal}{Phys. Rev. Lett.} \textbf{\bibinfo{volume}{81}},
  \bibinfo{pages}{34} (\bibinfo{year}{1998}), \eprint{hep-ph/9804359}.

\bibitem[{\citenamefont{Abdallah et~al.}(2003)}]{DELPHI:2003uqw}
\bibinfo{author}{\bibfnamefont{J.}~\bibnamefont{Abdallah}} \bibnamefont{et~al.}
  (\bibinfo{collaboration}{DELPHI}), \bibinfo{journal}{Eur. Phys. J. C}
  \textbf{\bibinfo{volume}{31}}, \bibinfo{pages}{421} (\bibinfo{year}{2003}),
  \eprint{hep-ex/0311019}.

\bibitem[{\citenamefont{Cynolter and Lendvai}(2008)}]{Cynolter:2008ea}
\bibinfo{author}{\bibfnamefont{G.}~\bibnamefont{Cynolter}} \bibnamefont{and}
  \bibinfo{author}{\bibfnamefont{E.}~\bibnamefont{Lendvai}},
  \bibinfo{journal}{Eur. Phys. J. C} \textbf{\bibinfo{volume}{58}},
  \bibinfo{pages}{463} (\bibinfo{year}{2008}), \eprint{0804.4080}.

\bibitem[{\citenamefont{Sirunyan et~al.}(2019)}]{CMS:2018yfx}
\bibinfo{author}{\bibfnamefont{A.~M.} \bibnamefont{Sirunyan}}
  \bibnamefont{et~al.} (\bibinfo{collaboration}{CMS}), \bibinfo{journal}{Phys.
  Lett. B} \textbf{\bibinfo{volume}{793}}, \bibinfo{pages}{520}
  (\bibinfo{year}{2019}), \eprint{1809.05937}.

\bibitem[{\citenamefont{Aalbers et~al.}(2022)}]{LZ:2022ufs}
\bibinfo{author}{\bibfnamefont{J.}~\bibnamefont{Aalbers}} \bibnamefont{et~al.}
  (\bibinfo{collaboration}{LZ}) (\bibinfo{year}{2022}), \eprint{2207.03764}.

\bibitem[{\citenamefont{Griest and Seckel}(1991)}]{Griest:1990kh}
\bibinfo{author}{\bibfnamefont{K.}~\bibnamefont{Griest}} \bibnamefont{and}
  \bibinfo{author}{\bibfnamefont{D.}~\bibnamefont{Seckel}},
  \bibinfo{journal}{Phys. Rev. D} \textbf{\bibinfo{volume}{43}},
  \bibinfo{pages}{3191} (\bibinfo{year}{1991}).

\bibitem[{\citenamefont{Edsjo and Gondolo}(1997)}]{Edsjo:1997bg}
\bibinfo{author}{\bibfnamefont{J.}~\bibnamefont{Edsjo}} \bibnamefont{and}
  \bibinfo{author}{\bibfnamefont{P.}~\bibnamefont{Gondolo}},
  \bibinfo{journal}{Phys. Rev. D} \textbf{\bibinfo{volume}{56}},
  \bibinfo{pages}{1879} (\bibinfo{year}{1997}), \eprint{hep-ph/9704361}.

\bibitem[{\citenamefont{B\'elanger et~al.}(2015)\citenamefont{B\'elanger,
  Boudjema, Pukhov, and Semenov}}]{Belanger:2014vza}
\bibinfo{author}{\bibfnamefont{G.}~\bibnamefont{B\'elanger}},
  \bibinfo{author}{\bibfnamefont{F.}~\bibnamefont{Boudjema}},
  \bibinfo{author}{\bibfnamefont{A.}~\bibnamefont{Pukhov}}, \bibnamefont{and}
  \bibinfo{author}{\bibfnamefont{A.}~\bibnamefont{Semenov}},
  \bibinfo{journal}{Comput. Phys. Commun.} \textbf{\bibinfo{volume}{192}},
  \bibinfo{pages}{322} (\bibinfo{year}{2015}), \eprint{1407.6129}.

\bibitem[{\citenamefont{Alloul et~al.}(2014)\citenamefont{Alloul, Christensen,
  Degrande, Duhr, and Fuks}}]{Alloul:2013bka}
\bibinfo{author}{\bibfnamefont{A.}~\bibnamefont{Alloul}},
  \bibinfo{author}{\bibfnamefont{N.~D.} \bibnamefont{Christensen}},
  \bibinfo{author}{\bibfnamefont{C.}~\bibnamefont{Degrande}},
  \bibinfo{author}{\bibfnamefont{C.}~\bibnamefont{Duhr}}, \bibnamefont{and}
  \bibinfo{author}{\bibfnamefont{B.}~\bibnamefont{Fuks}},
  \bibinfo{journal}{Comput. Phys. Commun.} \textbf{\bibinfo{volume}{185}},
  \bibinfo{pages}{2250} (\bibinfo{year}{2014}), \eprint{1310.1921}.

\bibitem[{\citenamefont{Bhattacharya
  et~al.}(2017{\natexlab{b}})\citenamefont{Bhattacharya, Poulose, and
  Ghosh}}]{Bhattacharya:2016ysw}
\bibinfo{author}{\bibfnamefont{S.}~\bibnamefont{Bhattacharya}},
  \bibinfo{author}{\bibfnamefont{P.}~\bibnamefont{Poulose}}, \bibnamefont{and}
  \bibinfo{author}{\bibfnamefont{P.}~\bibnamefont{Ghosh}},
  \bibinfo{journal}{JCAP} \textbf{\bibinfo{volume}{04}}, \bibinfo{pages}{043}
  (\bibinfo{year}{2017}{\natexlab{b}}), \eprint{1607.08461}.

\bibitem[{\citenamefont{Cirelli et~al.}(2006)\citenamefont{Cirelli, Fornengo,
  and Strumia}}]{Cirelli:2005uq}
\bibinfo{author}{\bibfnamefont{M.}~\bibnamefont{Cirelli}},
  \bibinfo{author}{\bibfnamefont{N.}~\bibnamefont{Fornengo}}, \bibnamefont{and}
  \bibinfo{author}{\bibfnamefont{A.}~\bibnamefont{Strumia}},
  \bibinfo{journal}{Nucl. Phys. B} \textbf{\bibinfo{volume}{753}},
  \bibinfo{pages}{178} (\bibinfo{year}{2006}), \eprint{hep-ph/0512090}.

\bibitem[{\citenamefont{Bhattacharya
  et~al.}(2019{\natexlab{b}})\citenamefont{Bhattacharya, Ghosh, and
  Sahu}}]{Bhattacharya:2018cgx}
\bibinfo{author}{\bibfnamefont{S.}~\bibnamefont{Bhattacharya}},
  \bibinfo{author}{\bibfnamefont{P.}~\bibnamefont{Ghosh}}, \bibnamefont{and}
  \bibinfo{author}{\bibfnamefont{N.}~\bibnamefont{Sahu}},
  \bibinfo{journal}{JHEP} \textbf{\bibinfo{volume}{02}}, \bibinfo{pages}{059}
  (\bibinfo{year}{2019}{\natexlab{b}}), \eprint{1809.07474}.

\bibitem[{\citenamefont{Feng et~al.}(2003)\citenamefont{Feng, Rajaraman, and
  Takayama}}]{Feng:2003uy}
\bibinfo{author}{\bibfnamefont{J.~L.} \bibnamefont{Feng}},
  \bibinfo{author}{\bibfnamefont{A.}~\bibnamefont{Rajaraman}},
  \bibnamefont{and} \bibinfo{author}{\bibfnamefont{F.}~\bibnamefont{Takayama}},
  \bibinfo{journal}{Phys. Rev. D} \textbf{\bibinfo{volume}{68}},
  \bibinfo{pages}{063504} (\bibinfo{year}{2003}), \eprint{hep-ph/0306024}.

\bibitem[{\citenamefont{Coy et~al.}(2021)\citenamefont{Coy, Gupta, and
  Hambye}}]{Coy:2021sse}
\bibinfo{author}{\bibfnamefont{R.}~\bibnamefont{Coy}},
  \bibinfo{author}{\bibfnamefont{A.}~\bibnamefont{Gupta}}, \bibnamefont{and}
  \bibinfo{author}{\bibfnamefont{T.}~\bibnamefont{Hambye}},
  \bibinfo{journal}{Phys. Rev. D} \textbf{\bibinfo{volume}{104}},
  \bibinfo{pages}{083024} (\bibinfo{year}{2021}), \eprint{2104.00042}.

\bibitem[{\citenamefont{Cheng and Li}(1980)}]{Cheng:1980qt}
\bibinfo{author}{\bibfnamefont{T.~P.} \bibnamefont{Cheng}} \bibnamefont{and}
  \bibinfo{author}{\bibfnamefont{L.-F.} \bibnamefont{Li}},
  \bibinfo{journal}{Phys. Rev. D} \textbf{\bibinfo{volume}{22}},
  \bibinfo{pages}{2860} (\bibinfo{year}{1980}).

\bibitem[{\citenamefont{Arhrib et~al.}(2011)\citenamefont{Arhrib, Benbrik,
  Chabab, Moultaka, Peyranere, Rahili, and Ramadan}}]{Arhrib:2011uy}
\bibinfo{author}{\bibfnamefont{A.}~\bibnamefont{Arhrib}},
  \bibinfo{author}{\bibfnamefont{R.}~\bibnamefont{Benbrik}},
  \bibinfo{author}{\bibfnamefont{M.}~\bibnamefont{Chabab}},
  \bibinfo{author}{\bibfnamefont{G.}~\bibnamefont{Moultaka}},
  \bibinfo{author}{\bibfnamefont{M.~C.} \bibnamefont{Peyranere}},
  \bibinfo{author}{\bibfnamefont{L.}~\bibnamefont{Rahili}}, \bibnamefont{and}
  \bibinfo{author}{\bibfnamefont{J.}~\bibnamefont{Ramadan}},
  \bibinfo{journal}{Phys. Rev. D} \textbf{\bibinfo{volume}{84}},
  \bibinfo{pages}{095005} (\bibinfo{year}{2011}), \eprint{1105.1925}.

\bibitem[{\citenamefont{Baldini et~al.}(2016)}]{MEG:2016leq}
\bibinfo{author}{\bibfnamefont{A.~M.} \bibnamefont{Baldini}}
  \bibnamefont{et~al.} (\bibinfo{collaboration}{MEG}), \bibinfo{journal}{Eur.
  Phys. J. C} \textbf{\bibinfo{volume}{76}}, \bibinfo{pages}{434}
  (\bibinfo{year}{2016}), \eprint{1605.05081}.

\bibitem[{\citenamefont{Essig}(2008)}]{Essig:2007az}
\bibinfo{author}{\bibfnamefont{R.}~\bibnamefont{Essig}},
  \bibinfo{journal}{Phys. Rev. D} \textbf{\bibinfo{volume}{78}},
  \bibinfo{pages}{015004} (\bibinfo{year}{2008}), \eprint{0710.1668}.

\bibitem[{\citenamefont{Tucker-Smith and Weiner}(2001)}]{Tucker-Smith:2001myb}
\bibinfo{author}{\bibfnamefont{D.}~\bibnamefont{Tucker-Smith}}
  \bibnamefont{and} \bibinfo{author}{\bibfnamefont{N.}~\bibnamefont{Weiner}},
  \bibinfo{journal}{Phys. Rev. D} \textbf{\bibinfo{volume}{64}},
  \bibinfo{pages}{043502} (\bibinfo{year}{2001}), \eprint{hep-ph/0101138}.

\bibitem[{\citenamefont{Barman et~al.}(2019{\natexlab{b}})\citenamefont{Barman,
  Bhattacharya, Ghosh, Kadam, and Sahu}}]{Barman:2019tuo}
\bibinfo{author}{\bibfnamefont{B.}~\bibnamefont{Barman}},
  \bibinfo{author}{\bibfnamefont{S.}~\bibnamefont{Bhattacharya}},
  \bibinfo{author}{\bibfnamefont{P.}~\bibnamefont{Ghosh}},
  \bibinfo{author}{\bibfnamefont{S.}~\bibnamefont{Kadam}}, \bibnamefont{and}
  \bibinfo{author}{\bibfnamefont{N.}~\bibnamefont{Sahu}},
  \bibinfo{journal}{Phys. Rev. D} \textbf{\bibinfo{volume}{100}},
  \bibinfo{pages}{015027} (\bibinfo{year}{2019}{\natexlab{b}}),
  \eprint{1902.01217}.

\bibitem[{\citenamefont{Hisano et~al.}(2011)\citenamefont{Hisano, Ishiwata,
  Nagata, and Takesako}}]{Hisano:2011cs}
\bibinfo{author}{\bibfnamefont{J.}~\bibnamefont{Hisano}},
  \bibinfo{author}{\bibfnamefont{K.}~\bibnamefont{Ishiwata}},
  \bibinfo{author}{\bibfnamefont{N.}~\bibnamefont{Nagata}}, \bibnamefont{and}
  \bibinfo{author}{\bibfnamefont{T.}~\bibnamefont{Takesako}},
  \bibinfo{journal}{JHEP} \textbf{\bibinfo{volume}{07}}, \bibinfo{pages}{005}
  (\bibinfo{year}{2011}), \eprint{1104.0228}.

\bibitem[{\citenamefont{Calibbi et~al.}(2018)\citenamefont{Calibbi,
  Lopez-Honorez, Lowette, and Mariotti}}]{Calibbi:2018fqf}
\bibinfo{author}{\bibfnamefont{L.}~\bibnamefont{Calibbi}},
  \bibinfo{author}{\bibfnamefont{L.}~\bibnamefont{Lopez-Honorez}},
  \bibinfo{author}{\bibfnamefont{S.}~\bibnamefont{Lowette}}, \bibnamefont{and}
  \bibinfo{author}{\bibfnamefont{A.}~\bibnamefont{Mariotti}},
  \bibinfo{journal}{JHEP} \textbf{\bibinfo{volume}{09}}, \bibinfo{pages}{037}
  (\bibinfo{year}{2018}), \eprint{1805.04423}.

\bibitem[{\citenamefont{Belyaev et~al.}(2021)\citenamefont{Belyaev, Prestel,
  Rojas-Abbate, and Zurita}}]{Belyaev:2020wok}
\bibinfo{author}{\bibfnamefont{A.}~\bibnamefont{Belyaev}},
  \bibinfo{author}{\bibfnamefont{S.}~\bibnamefont{Prestel}},
  \bibinfo{author}{\bibfnamefont{F.}~\bibnamefont{Rojas-Abbate}},
  \bibnamefont{and} \bibinfo{author}{\bibfnamefont{J.}~\bibnamefont{Zurita}},
  \bibinfo{journal}{Phys. Rev. D} \textbf{\bibinfo{volume}{103}},
  \bibinfo{pages}{095006} (\bibinfo{year}{2021}), \eprint{2008.08581}.

\bibitem[{\citenamefont{Ghosh et~al.}(2022{\natexlab{b}})\citenamefont{Ghosh,
  Ghosh, and Roy}}]{Ghosh:2022fzp}
\bibinfo{author}{\bibfnamefont{P.}~\bibnamefont{Ghosh}},
  \bibinfo{author}{\bibfnamefont{T.}~\bibnamefont{Ghosh}}, \bibnamefont{and}
  \bibinfo{author}{\bibfnamefont{S.}~\bibnamefont{Roy}}
  (\bibinfo{year}{2022}{\natexlab{b}}), \eprint{2211.15640}.

\bibitem[{\citenamefont{Akhmedov}(2014)}]{Akhmedov:2014kxa}
\bibinfo{author}{\bibfnamefont{E.}~\bibnamefont{Akhmedov}},
  \emph{\bibinfo{title}{{Majorana neutrinos and other Majorana particles:Theory
  and experiment}}} (\bibinfo{year}{2014}), \eprint{1412.3320}.

\bibitem[{\citenamefont{Hisano et~al.}(2010)\citenamefont{Hisano, Ishiwata, and
  Nagata}}]{Hisano:2010fy}
\bibinfo{author}{\bibfnamefont{J.}~\bibnamefont{Hisano}},
  \bibinfo{author}{\bibfnamefont{K.}~\bibnamefont{Ishiwata}}, \bibnamefont{and}
  \bibinfo{author}{\bibfnamefont{N.}~\bibnamefont{Nagata}},
  \bibinfo{journal}{Phys. Lett. B} \textbf{\bibinfo{volume}{690}},
  \bibinfo{pages}{311} (\bibinfo{year}{2010}), \eprint{1004.4090}.

\bibitem[{\citenamefont{Amintaheri}(2022)}]{Amintaheri:2022coc}
\bibinfo{author}{\bibfnamefont{R.}~\bibnamefont{Amintaheri}}
  (\bibinfo{year}{2022}), \eprint{2211.11899}.

\end{thebibliography}
\end{document}